\documentclass[journal,9pt]{IEEEtran}

\usepackage[font=scriptsize,caption=false,labelsep=space]{subfig}
\usepackage{mathrsfs}
\usepackage{graphicx,float,wrapfig,amsmath}
\usepackage[]{algorithmicx}
\usepackage{algpseudocode,algorithm}
\usepackage{balance}

\usepackage{amsmath}
\usepackage{amssymb}
\usepackage{amsthm}
\usepackage{graphicx}
\usepackage{epstopdf}
\usepackage{times}
\usepackage{textcomp,cite}
\usepackage{color}
\usepackage{url}
\usepackage{cuted}

\usepackage{multirow}
\usepackage[table]{xcolor}
\usepackage{threeparttable}

\DeclareMathOperator*{\argmax}{argmax} 
\DeclareMathOperator*{\maximize}{maximize} 
\DeclareMathOperator*{\minimize}{minimize} 
\DeclareMathOperator*{\subjectto}{subject \hspace{3pt} to:} 

\begin{document}
	
	\title{Joint Antenna Selection and Hybrid Beamformer Design using Unquantized and Quantized Deep Learning Networks}
	\author{Ahmet~M.~Elbir and Kumar Vijay~Mishra
		\thanks{A. M. E. is with the Department of Electrical and Electronics Engineering, Duzce University, Duzce, Turkey. E-mail: ahmetelbir@duzce.edu.tr.}
		\thanks{K. V. M. is with The University of Iowa, Iowa City, IA 52242 USA. E-mail: kumarvijay-mishra@uiowa.edu.}
	}
	\maketitle
	
	\begin{abstract}
		In millimeter wave communications, multiple-input-multiple-output (MIMO) systems use large antenna arrays to achieve high gain and spectral efficiency. These massive MIMO systems employ hybrid beamformers to reduce power consumption associated with fully digital beamforming in large arrays. Further savings in cost and power is possible through use of subarrays. Unlike prior works which resort to large latency methods such as optimization and greedy search for subarray selection, we propose a deep-learning-based approach in order to overcome the complexity issue without causing significant performance loss. We formulate antenna selection and hybrid beamformer design as a classification/prediction problem for convolutional neural networks (CNNs). For antenna selection, the CNN accepts the channel matrix as input and outputs a subarray with an optimal spectral efficiency. The resultant subarray channel matrix is then again fed to a CNN to obtain analog and baseband beamformers.  We train the CNNs with several noisy channel matrices that have different channel statistics in order to achieve a robust performance at the network output. Numerical experiments show that our CNN framework provides an order better spectral efficiency and is 10 times faster than the conventional techniques. Further investigations with quantized-CNNs show that the proposed network, saved in no more than 5 bits, is also suited for digital mobile devices.
	\end{abstract}
	
	
	\section{Introduction}
	\label{sec:Intro}	
	The conventional cellular communications systems suffer from spectrum shortage while the demand for wider bandwidth and higher data rates is continuously increasing \cite{mishra2019toward}. 
	In this context, millimeter wave (mm-Wave) band, formally defined with the frequency range 30-300 GHz, is a preferred candidate for fifth-generation (5G) communications technology 
	\cite{dokhanchi2019mmWave,mishra2019doppler,hodge2019reconfigurable,ayyar2019robust}. Compared to sub-6 GHz transmissions envisaged in 5G, the mm-Wave signals encounter a more complex propagation environment that is characterized by higher scattering, severe penetration losses, lower diffraction, and higher path loss for fixed transmitter and receiver gains 
	\cite{mishra2019toward}. These losses are compensated by providing beamforming power gain through massive number of antennas at both transmitter and receiver. Such a massive multiple-input-multiple-output (MIMO) structure \cite{mimoScalingUp,alaee2019discrete} enhances the signal-to-noise ratio (SNR) at the reception.
	
	The wide mm-Wave bandwidth enables higher data rates for communications. Since the Nyquist sampling rate is twice the baseband bandwidth, mm-Wave receivers require expensive, high-rate analog-to-digital converters (ADCs) \cite{mishra2018sub,hybridBFLowRes}. The power consumption of an ADC increases with the sampling rate, more so at high frequencies \cite{tsui2004digital}, for a given architecture. At baseband, each full-resolution ADC consumes $15$-$795$ mW at $36$ MHz-$1.8$ GHz bandwidths. In addition, power consumed by other RF elements such as power amplifiers and data interface circuits in conjunction with large arrays renders it infeasible to utilize a separate radio- and intermediate-frequency (RF-IF) chain for each element. 
	To reduce these cost-power-hardware overheads and yet provide reasonable performance, hybrid beamforming architectures have been proposed for massive MIMO. Here, the signal is processed by both analog and digital beamformers \cite{mimoHybridLeus1,mimoHybridLeus2,mimoHybridLeus3,mimoRHeath,hybridBFAltMin}. 
	
	In the analog processing section of hybrid systems, it is common to employ phase shifters with constant modulus. Using analog switches, which are much simpler and cheaper than the phase shifters, it is possible to further make the overall system more energy-efficient by using subarrays of the larger full antenna array 
	\cite{mishra2018cognitive,na2018tendsur}. Optimal selection of subarray elements reduces the power consumption of the analog phase shifters and low-noise amplifiers (LNAs) \cite{mimoJointHBASTx,mimoJonitTransciveAntennaSelection,mimoJonitTransciveAntennaSelection2}. Very recent works consider the problem of antenna selection jointly with hybrid beamformer design to optimally trade-off cost and power efficiency \cite{mimoJonitTransciveAntennaSelection,mimoJonitTransciveAntennaSelection2,mimoJointHBAS,mimoJointHBAS2}. In particular, \cite{mimoJointHBAS} proposed antenna selection and analog precoder design with low-resolution phase shifters for multiple-input-single-output (MISO) systems. The hybrid beamformer designs suggested in \cite{mimoJonitTransciveAntennaSelection} and \cite{mimoJonitTransciveAntennaSelection2} involve a sub-optimum antenna selection strategy through quadratic approximation with smooth optimization. Similarly, the massive MIMO architecture in \cite{mimoJointHBAS2} employs greedy search for choosing sub-optimal subarrays.  
	
	Nearly all of these works provide sub-optimum solutions despite attempting various antenna selection criteria and optimization strategies. 
	Even while using branch-and-bound (BAB) algorithms - which provide good estimation of the lower and upper bounds of regions/branches of the search space in polynomial time - obtaining an optimum solution for massive MIMO subarray selection requires high computational burden \cite{mimoAntennaSelectionBAB}. In this paper,  to reduce the complexity (in cases where the optimum solution can still be obtained), we introduce an approach based on deep learning (DL) to find an optimum subarray jointly with the design of hybrid beamformers; the optimality is in the sense of achieving maximum spectral efficiency. 
	
	As a class of machine learning techniques, DL methods have gained much interest recently for solving many challenging problems such as visual object recognition \cite{deepLearningScience}, rainfall estimation \cite{mishra2018deep}, and language processing \cite{deppLearningRepresetation}. 
	These techniques offer advantages such as low computational complexity while solving optimization-based or combinatorial search problems as well as the ability to extrapolate new features from a limited set of features contained in a training set \cite{deepLearningScience}. Very recently, DL has received significant attention in addressing problems in communications signal processing such as channel estimation \cite{mimoDLChannelEstimation}, 
	direction-of-arrival (DoA) estimation \cite{mimoDLHybrid} 
	analog beam selection \cite{hodge2019joint,hodge2019rf,hodge2019multi} and beam management in dense mm-Wave networks \cite{deepLearning4densemmWave}. At the physical layer of wireless communications, DL has been applied for signal detection \cite{mimoDLDetection} and channel estimation \cite{mimoDLOverAir}. An end-to-end single-input-single-output (SISO) communications scenario is modeled in \cite{mimoDLOverAir} and \cite{mimoDLOverAir2} by using auto-encoders. In \cite{mimoDLCSIFeedBack}, auto-encoders are employed for channel state information (CSI) feedback.  A sub-optimum method based on support vector machines (SVMs) is proposed in \cite{mimoDLHybrid} for selecting analog beamforming vector. In a recent work \cite{mimoDeepPrecoderDesign},  multilayer perceptrons (MLP) are used for precoder design in a single-user mm-Wave scenario.
	
	In this paper, we exploit DL to simultaneously select antenna elements and design hybrid beamformer. This joint problem as well as the stand-alone hybrid beamforming remain unexamined in the previous DL works. Specifically, we design a convolutional neural network (CNN) to achieve both tasks sequentially. 
	The element selection problem is cast as a classification problem~\cite{elbirAPS2019}. A similar DL approach was adopted for radar antenna arrays recently in \cite{deepLearningAntennaSelectionElbir}. 
	We further incorporate the hybrid beamformer design in this DL framework by exploiting the structure of analog beamformers  which are obtained by minimizing the cost between hybrid and unconstrained beamformers. The optimization problem is cast jointly with the antenna selection problem and it is solved  by MATLAB-based Manopt algorithm \cite{manopt} via manifold optimization (MO) \cite{hybridBFAltMin}.
	
	In our formulation, a CNN accepts channel matrix as input and provides the subarray that maximizes the spectral efficiency. Once the antenna selection is finalized, the corresponding partial channel matrix is fed to a second CNN which then chooses the best RF beamformer and constructs the corresponding baseband beamformer. To train both CNN models, different realizations of the channel matrix are used and the input data are labeled by the selected subarray/RF chains with the highest spectral efficiency. Even though our proposed network structures require channel matrix as an input, precise knowledge of this matrix is  only required in the training stage to obtain the labels of the network. In the prediction stage, where the RF beamformers are estimated, precise channel knowledge is not necessary. Both CNNs are trained with channel matrices generated for different user location, channel gains and number of user clusters. Furthermore, each realization of channel matrix in the training data is corrupted by synthetic noise so that the performance of the learning network does not deteriorate with noisy test inputs. 
	
	We evaluate the performance of the proposed framework over several experiments and show that proposed CNN approach provides significantly better performance as compared to the  optimization and greedy based techniques \cite{mimoRHeath,hybridBFLowRes,mimoDeepPrecoderDesign,sohrabiNarrowband}.
	In order to account for time-varying channel and user parameters, we use several channel realizations with added noise. We train the networks with huge training data (\texttildelow240000 input samples) with noisy channel matrices. As a result, the classification accuracy quickly reaches $100\%$ wherein optimum antenna selection and RF beamformer design are accomplished. 
	The CNN is trained offline and, hence, all the computational overhead is taken into account for data generation and training. The classification and prediction time for our proposed approach is at least 10 times faster than the conventional antenna selection techniques as well as hybrid beamformer design algorithms. 
	
	Finally, our approach is helpful in reducing the computational burden involved in hybrid beamformers by simply feeding the channel matrix to the network. This requires using CNNs in mobile devices where the data are collected in digital form. Since existing deep neural network models are computationally and memory intensive, they cannot be deployed in devices with low memory resources and low overhead requirements. These constraints have driven investigation into compression of deep neural networks. One of the common approaches is to quantize the CNN weights \cite{elbir2019robust}, \cite{quantizedCNN}. In this paper, we investigate the performance of the proposed framework when the weights of the CNNs are quantized. While quantized-CNN structures are recently studied for image classification purposes, ours is the first work that examines quantized-CNNs for communications. Preliminary results of our work appeared in \cite{elbirAPS2019} and \cite{elbir2019robust}; while a basic formulation suggested in \cite{elbirAPS2019} solved the joint problem for a specific hybrid beamforming scheme, \cite{elbir2019robust} proposed a quantized-CNN approach for the same scheme but did not include antenna selection.

	Throughout the paper, we denote the identity matrix of size $N\times N$ as $\mathbf{I}_N$. $(\cdot)^T$ and $(\cdot)^H$ denote transpose and  the conjugate transpose operations, respectively. For a matrix $\mathbf{A}$ and a vector $\mathbf{a}$, $[\mathbf{A}]_{:,i}$ and $[\mathbf{A}]_{i,j}$ denote the $i$th column and $(i,j)$th element of matrix $\mathbf{A}$, $[\mathbf{a}]_i$ means the $i$th element of vector $\mathbf{a}$, respectively. The notation $|\mathbf{A}|$ denotes the determinant of matrix $\mathbf{A}$ whereas $|a|$ is the absolute value of the scalar $a$. The function $\mathbb{E}\{\cdot\}$ provides the statistical expectation of its argument and  $\angle \{\cdot\}$ measures the angle of complex quantity.

	\begin{figure*}
		\centering
		{\includegraphics[width=0.8\textwidth] 
			{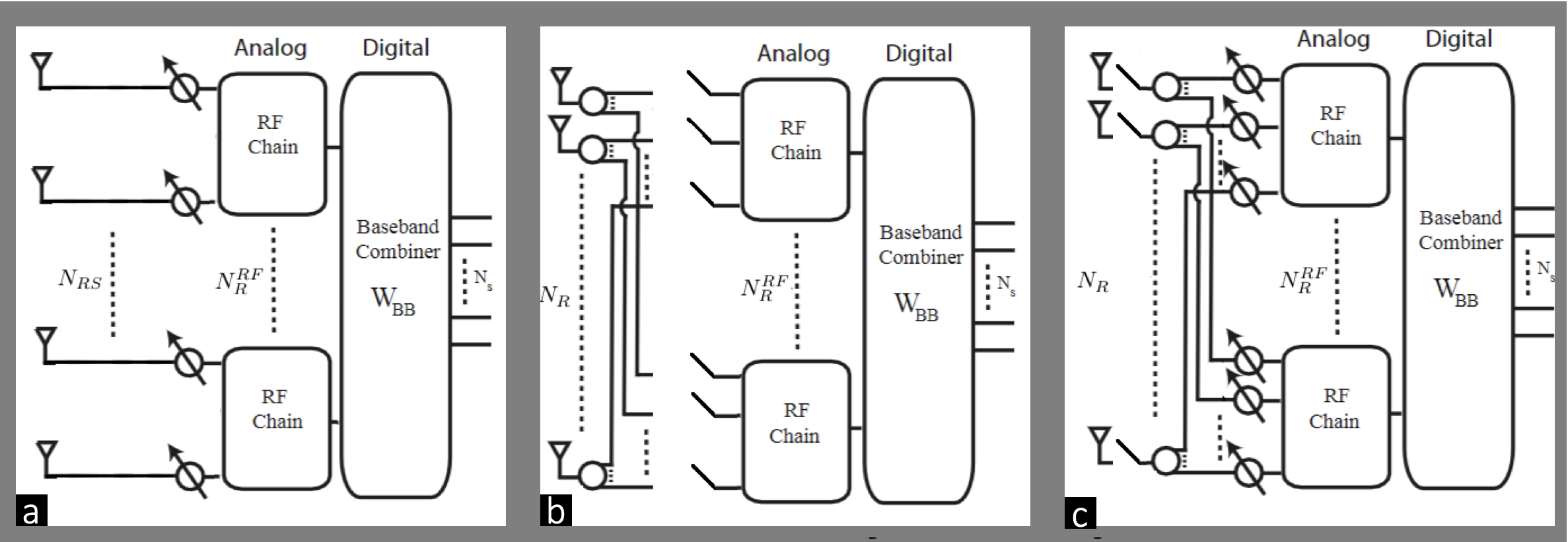} }
		\caption{Receiver architectures with antenna selection for single user mmWave MIMO systems. (a) Scheme 1: Fixed subarray with  fully-connected phase shifters. (b) Scheme 2: Switching network without optimized antenna selection and no phase shifters. 
			(c) Scheme 3:	Switching network with optimized antenna selection and phase shifters.
		}
		
		\label{figAntennaSelectionSchemes}
	\end{figure*}
	
	\section{System Model For mm-Wave MIMO Systems}
	\label{sec:SystemModel}
	Consider a single user mm-Wave MIMO system with 
	$N_T$ and $N_R$ transmit and receive antennas, respectively (Fig.~\ref{figAntennaSelectionSchemes}). Assume that $N_S$ data streams are desired to be transmitted to the receiver where the antenna selection is performed to select a subarray with $N_{RS}$ antennas out of $N_R$. 
	There are $N_T^{RF}$ and $N_R^{RF}$ RF beamformers at transmit and receive sides such that $N_S \leq N_{T}^{RF} \leq N_T$ and $N_S\leq N_R^{RF}\leq N_{RS} \leq N_R$. The hybrid precoder structure applies the baseband precoder $\mathbf{F}_{BB}\in \mathbb{C}^{N_T^{RF}\times N_S}$ to the transmit signal vector $\mathbf{s}\in \mathbb{C}^{N_S}$, where $\mathbb{E}\{\mathbf{s}\mathbf{s}^H\} = \mathbf{I}_{N_S}/N_S$. Then, the signal is passed through RF precoders $\mathbf{F}_{RF}\in \mathbb{C}^{N_T\times N_T^{RF}}$ (constructed using phase shifters) to $N_T$ transmit antennas. The RF precoder has equal-norm elements so that $[[\mathbf{F}_{RF}]_{:,i} [\mathbf{F}_{RF}]_{:,i}^H ]_{i,i}=1/N_T$. The power of the transmitter is constrained to $||\mathbf{F}_{RF}\mathbf{F}_{BB} ||_\mathcal{F}= N_S$.  The transmitted signal at RF stage is $\mathbf{x} = \mathbf{F}_{RF} \mathbf{F}_{BB} \mathbf{s} \in \mathbb{C}^{N_T}$. Assuming a narrowband block-fading channel, the received signal at $N_R$ antennas is \cite{mmWaveModel1,mimoRHeath}
	\begin{align}
	\label{def:ReceivedSignalAllAntennas}
	\mathbf{y}^{\text{Full}} = \sqrt{\rho} \mathbf{H}\mathbf{F}_{RF}\mathbf{F}_{BB}\mathbf{s} + \mathbf{n},
	\end{align}
	where $\mathbf{y}^{\text{Full}}\in \mathbb{C}^{N_R}$ is the output of $N_R$ antennas at the receiver, $\rho$ is average received power, $\mathbf{n}\in \mathbb{C}^{N_R}$ is the additive white Gaussian noise (AWGN) with $\mathbf{n} \sim \mathcal{CN}(\mathbf{0}, \sigma_n^2 \mathbf{I}_{N_R})$, and $\mathbf{H}\in \mathbb{C}^{N_R \times N_T}$ is the channel matrix with $||\mathbf{H}||_\mathcal{F} = N_RN_T$.

	A standard analog-digital hybrid beamformer operates on a full array. When a subarray is employed, then beamformers must be derived for reduced dimensions. We model the antenna selection problem to select the \textit{best} antenna subset from all possible antenna subarrays. Throughout this paper, we use the terms ``subarray'' and ``subset'' interchangeably; a subarray being a subset of indices (or antenna positions) of a full array configuration. 
	A popular scheme is to employ 
	a predetermined subarray with $N_{RS}$ antennas is selected from a full array of $N_R$ elements (Fig.~\ref{figAntennaSelectionSchemes}a). Each subarray feeds into a fully-connected phase shifter network of size $N_R^{RF}$ with a single RF chain. This has the complexity of phase shifters but the antenna selection process is not optimized. 
	Another common receiver architecture feeds the antennas directly to the RF chains thereby eliminating the phase shifters completely. 
	Here, 
	each RF chain is connected to the $N_{R}$ antennas of which $N_{RS}$ elements are selected using switches (Fig.~\ref{figAntennaSelectionSchemes}b).  In this case, the entries of the combiner matrix are either 1 or 0 to indicate the selected or unselected antennas, respectively. This is the simplest structure with no phase shifters. However, the antenna selection is not optimzed and the elements are determined by simply choosing the largest absolute values in each column of the
	channel matrix. 
	Finally, Fig.~\ref{figAntennaSelectionSchemes}c shows a receiver that employs a switching network with phase shifters. In this system, 
	a subarray with $N_{RS}$ antennas is selected from a full array comprising $N_{R}$ antennas. The subarray is connected to a phase shifter network of size $N_R^{RF}$ which may apply an optimization procedure for antenna selection to achieve greater efficiency.

	
	Our mm-Wave channel representation is based on the Saleh-Valenzuela (SV) model that utilizes the clustered channel model \cite{mimoChannelModel1,mimoChannelModel2}. Here, the channel matrix $\mathbf{H}$ includes the contributions of $N_{c}$ scattering  clusters, each of which has $N_{\text{ray}}$ paths. We have
	\begin{align}
	\mathbf{H} = \gamma \sum_{i,j} \alpha_{ij} g_R(\Theta_R^{(ij)}) g_T(\Theta_T^{(ij)}) \mathbf{a}_R(\Theta_R^{(ij)}) \mathbf{a}_T^H(\Theta_T^{(ij)}),
	\end{align} 
	where  $\Theta_R^{(ij)}=(\phi_R^{(ij)}, \theta_R^{(ij)})$ and $\Theta_T^{(ij)}=(\phi_T^{(ij)}, \theta_T^{(ij)})$, respectively, denote the angle of arrivals and angle of departures wherein the azimuth (elevation) angle is denoted by $\phi$ ($\theta$), $\gamma = \sqrt{ N_T N_{RS}/(N_\mathrm{c}N_{\mathrm{ray}})}$ is the normalization factor, and $\alpha_{ij}$ is the complex channel gain associated with the $i$th scattering cluster and $j$th path for $i = 1,\dots,N_\mathrm{c}$ and $j = 1,\dots, N_\mathrm{ray}$. The antenna element gains for receive and transmit antennas are $g_R(\Theta_R^{(ij)})$ and $ g_T(\Theta_T^{(ij)})$, respectively. The steering vector representing the array response at the transmitter (receiver) is $\mathbf{a}_T(\Theta_T^{(ij)}) \in N_T\times 1$ ($\mathbf{a}_R(\Theta_R^{(ij)}) \in N_R \times 1$). The $n$th element of $\mathbf{a}_R(\Theta_R^{(ij)})$ is 
	\begin{align}
	[\mathbf{a}_R(\Theta_R^{(ij)})]_n = \exp\left\{ -\frac{2\pi}{\lambda}\mathbf{p}_n^T \mathbf{r}(\Theta_R ^{(ij)}) \right\},
	\end{align}
	where $\mathbf{p}_n =[x_n,y_n,z_n]^T$ is the position of the $n$th antenna in Cartesian coordinate system and $\mathbf{r}(\Theta_R^{(ij)}) = [\sin(\phi_R^{(ij)})\cos(\theta_R^{(ij)}),\sin(\phi_R^{(ij)})\sin(\theta_R^{(ij)}),\cos(\theta_R^{(ij)})]^T$. The transmit steering vector $\mathbf{a}_T(\Theta_T^{(ij)})$ is defined similarly.

	In practice, the estimation process of the channel matrix is a challenging task, especially in case of a large number of antennas in massive MIMO systems \cite{channelEstLargeArrays,channelEstimation1}. Additionally, the mm-Wave channel has short coherence times \cite{coherenceTimeRef,mishra2019toward}. In practice, the estimated channel matrix could be obtained via one of the several channel estimation techniques \cite{mimoChannelModel2,channelEstimation1CS,channelEstimation1,mimoAngleDomainFaiFai,mimoHybridLeus2}. For robust performance against imperfect channel estimates, our proposed DL framework feeds the deep network with several channel realizations which are corrupted by synthetic noise in the training stage. This is an offline process. Note that the perfect knowledge of channel matrix ${\mathbf{H}}$ is only required{\footnote{{In order to achieve antenna selection and train the deep network, full channel information of size $N_\mathrm{R}\times N_\mathrm{T}$ is required. This is also a common requirement in most antenna selection algorithms \cite{mimoJonitTransciveAntennaSelection,mimoJonitTransciveAntennaSelection2,antennaSelectionViaCO,mimoAntennaSelectionBAB,greedyAntennaSelection}. After antenna selection, the hybrid beamforming network requires only the partial channel matrix of size $N_\mathrm{RS}\times N_\mathrm{T}$ corresponding to the selected antennas.}}} at the training stage to obtain the labels (e.g., hybrid beamformer matrices). During the testing stage when the network predicts the beamformer weights, the network does not necessarily require the perfect CSI. Our numerical experiments demonstrate that the proposed approach can handle the corrupted channel matrix case and exhibits satisfactory performance regarding the achievable spectral efficiency.

	In the hybrid beamformer, analog and digital beamformers are obtained to maximize the spectral efficiency. Often, this is achieved by exploiting the structure of the mm-Wave channel matrix \cite{mimoAngleDomainFaiFai}. Using the full antenna array at the receiver, the received signal in (\ref{def:ReceivedSignalAllAntennas}) is processed by analog and baseband combiners to yield
	\begin{align}
	\label{def:receivedSignalFullArray}
	\bar{\mathbf{y}} &= \mathbf{W}_{BB}^{\text{Full}^H} \mathbf{W}_{RF}^{\text{Full}^H} \mathbf{y}^{\text{Full}} \nonumber \\
	&= \sqrt{\rho}\mathbf{W}_{BB}^{\text{Full}^H} \mathbf{W}_{RF}^{\text{Full}^H} \mathbf{H}\mathbf{F}_{RF}\mathbf{F}_{BB}\mathbf{s} + \mathbf{W}_{BB}^{\text{Full}^H} \mathbf{W}_{RF}^{\text{Full}^H}\mathbf{n},
	\end{align}
	where  $\mathbf{W}_{RF}^{\text{Full}}\in \mathbb{C}^{N_{R} \times N_R^{RF}}$ is the analog combiner with the constrained $[[\mathbf{W}_{RF}^{\text{Full}}]_{:,i} [\mathbf{W}_{RF}^{\text{Full}}]_{:,i}^H ]_{i,i}=1/N_{R}$ and  $\mathbf{W}_{BB}^{\text{Full}}\in \mathbb{C}^{N_R^{RF}\times N_S}$ denotes the baseband combiner matrix. Assuming that the Gaussian symbols are transmitted through the mm-Wave channel, we define the spectral efficiency \cite{mimoRHeath,mimoHybridLeus1,mimoHybridLeus2,mimoHybridLeus3} achieved from the full array as
	\par\noindent\small
	\begin{align}
	\label{def:RFull}
	&R^{\text{Full}} = \log_2 \bigg|\mathbf{I}_{N_S} + \frac{\rho}{N_S} \boldsymbol{\Lambda}_n^{\text{Full}^{-1}} \mathbf{W}_{BB}^{\text{Full}^H} \mathbf{W}_{RF}^{\text{Full}^H} \mathbf{H} \nonumber \\
	&\;\;\;\;\;\;\;\;\;\;\;\;\times \mathbf{F}_{RF}\mathbf{F}_{BB}   \mathbf{F}_{BB}^H\mathbf{F}_{RF}^H\mathbf{H}^H\mathbf{W}_{RF}^{\text{Full}}\mathbf{W}_{BB}^{\text{Full}} \bigg| ,
	\end{align}\normalsize
	\vspace*{4pt}
	\noindent 
	where $\boldsymbol{\Lambda}_n^{\text{Full}} = \sigma_n^2\mathbf{W}_{BB}^{\text{Full}^H} \mathbf{W}_{RF}^{\text{Full}^H} \mathbf{W}_{RF}^{\text{Full}} \mathbf{W}_{BB}^{\text{Full}} \in \mathbb{C}^{N_S\times N_S} $ is the covariance matrix of the noise term in (\ref{def:receivedSignalFullArray}) after analog combining. We now formulate the problem for subarray selection and obtaining the corresponding analog-digital beamformer in the following section.
	

	\section{Joint Antenna and RF Chain Selection} 
	\label{sec:AntennaRFSelection}
	Among the antenna selection schemes presented in the previous section, we focus on the Scheme 3 because this architecture requires optimization (the remaining configurations consider selecting a fixed subarray with/without phase shifters). In particular, our goal is to select the outputs of $N_{RS}$ antennas from the full array output $\mathbf{y}^{\text{Full}}$. 	Consequently, this also requires designing transmit and receive analog and baseband beamformers $\mathbf{F}_{RF}\in \mathbb{C}^{N_{T} \times N_T^{RF}}$, $\mathbf{W}_{RF}\in \mathbb{C}^{N_{RS} \times N_R^{RF}}$ and  $\mathbf{F}_{BB}\in \mathbb{C}^{N_{T}^{RF} \times N_{S}}$, $\mathbf{W}_{BB}\in \mathbb{C}^{N_{R}^{RF} \times N_{S}} $. In other words, the solution of joint antenna selection and hybrid beamformer design satisfies
	\par\noindent\small
	\begin{align}
	& \hspace{-2pt}\underset{\mathbf{Q},{\mathbf{F}}_{RF},{\mathbf{F}}_{BB},{\mathbf{W}}_{RF},{\mathbf{W}}_{BB}}{\maximize} \hspace{-10pt} \log_2 \bigg|\mathbf{I}_{N_S} +    \frac{\rho}{N_S\sigma_n^2} \big( \mathbf{W}_{BB}^H\mathbf{W}_{RF}^H  \mathbf{W}_{RF}\mathbf{W}_{BB}\big)^{-1} \nonumber\\ 
	&\mathbf{W}_{BB}^H\mathbf{W}_{RF}^H  \mathbf{H}_{\mathrm{sub}}
	\mathbf{F}_{RF}\mathbf{F}_{BB}  \mathbf{F}_{BB}^H\mathbf{F}_{RF}^H\mathbf{H}_{\mathrm{sub}}^H \mathbf{W}_{RF}\mathbf{W}_{BB} \bigg| \nonumber\\
	& \;\;\;\;\;\;\;\; \subjectto \;\;\;\;\;\;\; \mathbf{F}_{RF} \in \mathcal{F}_{RF},
	\;\;
	||\mathbf{F}_{RF}\mathbf{F}_{BB}||_\mathcal{F}^2 = N_S, \nonumber\\
	& \;\;\;\;\;\;\;\;\;\;\;\;\;\;\;\;\;\;\;\;\;\;\;\;\;\;\;\;\;\;\;\;	\mathbf{W}_{RF} \in \mathcal{W}_{RF},
	\;\;
	|| \mathrm{diag}\{\mathbf{Q}\}||_0 = N_{RS},
	\label{def:JointProblem}
	\end{align}\normalsize
	where $\mathcal{F}_{RF}$ and $\mathcal{W}_{RF}$ denote the feasible sets of analog beamformers, $\mathbf{H}_{\mathrm{sub}} = \mathbf{Q}\mathbf{H}$ is $N_{RS}\times N_T$ channel matrix of the selected antennas, and $\mathbf{Q}$ is the $N_{RS}\times N_R$ selection matrix whose $(i,j)$th entry is either 1 or 0. Even without antenna selection, the problem in (\ref{def:JointProblem}) is difficult to solve because of several matrix variables $\mathbf{Q}$, $\mathbf{F}_{RF}$, $\mathbf{W}_{RF}$ and  $\mathbf{F}_{BB}$, $\mathbf{W}_{BB}$   \cite{mimoOptProblem,mimoAntennaSelectionBAB}. Since obtaining a solution to (\ref{def:JointProblem}) in real-time is deemed infeasible, we propose a deep learning approach here 
	to achieve an optimum solution with less computational complexity. 
	We first cast the antenna selection stage as a classification problem as follows. 
	\subsection{Antenna Selection}
	\label{sec:AntennaSelection}
	In subarray selection, we are interested in picking $N_{RS}$ out of $N_R$ elements. This yields ${Q_A} = \left(\begin{array}{c}
	N_R \\ N_{RS}
	\end{array} \right)$ possible solutions. Therefore, choosing subarrays can be viewed as a classification problem with ${Q_A}$ classes. We define $\mathbb{S}$ as the set of all possible antenna subarray configurations, i.e., $\mathbb{S} = \{\mathbb{S}_1, \mathbb{S}_2,\dots,\mathbb{S}_{Q_A}  \},$
	where $\mathbb{S}_{q_A} = \{\mathbf{p}_1^{q_A},\mathbf{p}_2^{q_A},\dots,\mathbf{p}_{N_{RS}}^{q_A}  \}$ includes the antenna positions of the ${q_A}$th subarray configuration with $q_A \in \mathbb{Q}_A = \{1,\dots, Q_A\}$. Let $\mathbf{y}_{{q_A}}$ be an $N_{RS}\times 1$ vector containing the output signal of the selected antennas for the $q_A$th subarray configuration of the full array output $\mathbf{y}^{\text{Full}}$ with positions $\mathbb{S}_{q_A}$. Then, 
	\begin{align}
	\label{def:arrayOutputSelected}
	\mathbf{y}_{q_A} = \sqrt{\rho} \mathbf{H}_{{q_A}}\mathbf{F}_{RF}\mathbf{F}_{BB}\mathbf{s} + \mathbf{n}_{{q_A}},
	\end{align}
	where $\mathbf{H}_{{q_A}}$ is the $N_{RS}\times N_T$ channel matrix with selected antennas and $\mathbf{n}_{q_A}$ is similarly defined. At the receiver, analog and the baseband combiners - $\mathbf{W}_{RF}\in \mathbb{C}^{N_{RS} \times N_R^{RF}}$ and $\mathbf{W}_{BB}\in \mathbb{C}^{N_R^{RF}\times N_S}$, respectively - are applied to the received signal to produce the $N_S\times 1$ discrete-time signal $\bar{\mathbf{y}}_{q_A} = \mathbf{W}_{BB}^H \mathbf{W}_{RF}^H \mathbf{y}_{{q_A}}$ as
	\begin{align}
	\label{def:receivedSignal}
	\bar{\mathbf{y}}_{q_A} = \sqrt{\rho}\mathbf{W}_{BB}^H \mathbf{W}_{RF}^H \mathbf{H}_{q_A}\mathbf{F}_{RF}\mathbf{F}_{BB}\mathbf{s} + \mathbf{W}_{BB}^H \mathbf{W}_{RF}^H\mathbf{n}_{{q_A}}.
	\end{align}
	
	When the ${q_A}$th subarray is selected, the spectral efficiency \cite{capacityComputation}  of the mm-Wave channel is 
	\par\noindent\small
	\begin{align}
	\label{def:R}
	R({q_A}) = \log_2  \bigg|\mathbf{I}_{N_S} + \frac{\rho}{N_S} \boldsymbol{\Lambda}_n^{-1} \mathbf{W}_{BB}^H \mathbf{W}_{RF}^H \mathbf{H}_{q_A} \nonumber \\ \times \mathbf{F}_{RF}\mathbf{F}_{BB}  \mathbf{F}_{BB}^H\mathbf{F}_{RF}^H\mathbf{H}_{q_A}^H\mathbf{W}_{RF}\mathbf{W}_{BB} \bigg| ,
	\end{align}\normalsize
	where $\boldsymbol{\Lambda}_n = \sigma_n^2\mathbf{W}_{BB}^H \mathbf{W}_{RF}^H \mathbf{W}_{RF} \mathbf{W}_{BB} \in \mathbb{C}^{N_S\times N_S} $ corresponds to the noise term of the subarray output in (\ref{def:arrayOutputSelected}). Note that $R({q_A})$ depends on $q_A$ through $ \mathbf{H}_{q_A}$. 	By maximizing the spectral efficiency in (\ref{def:R}) over all subarray configurations, the best antenna subarray is obtained as
	\begin{align}
	\label{def:optProblem1}
	\bar{q}_A = \argmax_{q_A \in \mathbb{Q}_A} R(q_A),
	\end{align}
	where $\bar{q}_A$ denotes the subarray index with antenna positions $\mathbb{S}_{\bar{q}_A}$ which provide the maximum spectral efficiency. 
	
	While the optimization problem in (\ref{def:optProblem1}) yields the best subarray configuration, it does not impose any constraint on the hybrid beamformers. Here the problem in (\ref{def:optProblem1}) is cast by using unconstrained beamformers $\mathbf{F}_{q_A}^{\mathrm{opt}}\in \mathbb{C}^{N_T \times N_S}$ and $\mathbf{W}_{q_A}^{\mathrm{opt}}\in \mathbb{C}^{N_{RS}\times N_S }$. These can be obtained from the singular value decomposition (SVD) of the ${N_{RS} \times N_T }$ complex-valued channel matrix: $\mathbf{H}_{q_A} = \mathbf{U}_{q_A} \boldsymbol{\Sigma}_{q_A} \mathbf{V}_{q_A}^H$, where $\mathbf{U}_{q_A}\in \mathbb{C}^{N_{RS}\times \mathrm{rank}(\mathbf{H}_{q_A})}$ and $\mathbf{V}_{q_A}\in \mathbb{C}^{N_T \times \mathrm{rank}(\mathbf{H}_{q_A})}$ are the left and the right singular value matrices of the ${q_A}$th channel matrix, respectively, 
	and $\boldsymbol{\Sigma}_{q_A}$ is $\mathrm{rank}(\mathbf{H}_{q_A})\times \mathrm{rank}(\mathbf{H}_{q_A})$ matrix composed of the singular values of $\mathbf{H}_{q_A}$ in descending order. By decomposing $\boldsymbol{\Sigma}_{q_A}$ and $\mathbf{V}_{q_A}$ as $\boldsymbol{\Sigma}_{q_A} = \text{diag}\{ \boldsymbol{\Sigma}_{q_A}^{(1)},\boldsymbol{\Sigma}_{{q_A}}^{(2)}  \},\hspace{5pt} \mathbf{V}_{q_A} = [\mathbf{V}_{{q_A}}^{(1)},\mathbf{V}_{{q_A}}^{(2)}],$
	where $\mathbf{V}_{{q_A}}^{(1)}\in \mathbb{C}^{N_T\times N_S}$ and $\mathbf{V}_{{q_A}}^{(2)}\in \mathbb{C}^{N_T\times N_{RS}-N_S}$, one can readily select the unconstrained precoder as $\mathbf{F}_{q_A}^{\text{opt}} = \mathbf{V}_{{q_A}}^{(1)}$ \cite{mimoRHeath}. Using the unconstrained beamformer $\mathbf{F}_{q_A}^{\mathrm{opt}}$, $\mathbf{W}_{q_A}^{\mathrm{opt}}$ is computed as \cite{WoptCombiner}
	\begin{align}
	\mathbf{W}_{q_A}^{\mathrm{opt}} = \bigg( \frac{1}{\rho}\big( \mathbf{F}_{q_A}^{\mathrm{opt}^H}\mathbf{H}_{q_A}^H\mathbf{H}_{q_A}\mathbf{F}_{q_A}^{\mathrm{opt}}  + \frac{N_S\sigma_n^2}{\rho}\mathbf{I}_{N_S} \big)^{-1}  \mathbf{F}_{q_A}^{\mathrm{opt}^H}\mathbf{H}_{q_A}^H\bigg)^H. \nonumber
	\end{align}
	Using $\mathbf{F}_{q_A}^{\mathrm{opt}}$ and $\mathbf{W}_{q_A}^{\mathrm{opt}}$, the following  problem can be written, i.e.,
	\par\noindent\small
	\begin{align}
	& \underset{q_A \in \mathbb{Q}_A}{\maximize}\; \log_2 \bigg|\mathbf{I}_{N_S} +    \frac{\rho}{N_S\sigma_n^2} \big( \mathbf{W}_{q_A}^{\mathrm{opt}^H} \mathbf{W}_{q_A}^{\mathrm{opt}}\big)^{-1} \mathbf{W}_{q_A}^{\mathrm{opt}^H}  \mathbf{H}_{q_A}    \nonumber \\
	&\;\;\;\;\;\;\;\;\;\;\;\;\;\;\;\;\;\;\;\;\times \mathbf{F}_{q_A}^{\mathrm{opt}} \mathbf{F}_{q_A}^{\mathrm{opt}^H}\mathbf{H}_{q_A}^H \mathbf{W}_{q_A}^{\mathrm{opt}} \bigg|.
	\label{def:RAntennaSelection}
	\end{align}\normalsize

	The optimization problem in (\ref{def:RAntennaSelection}) uses  unconstrained beamformers $\mathbf{F}^{\mathrm{opt}}$, $\mathbf{W}^{\mathrm{opt}}$ for antenna selection and hybrid beamformer design. However, the "best" subarray obtained from  (\ref{def:RAntennaSelection}) would be different if  hybrid beamformers are used in the problem.
	Hence, we consider the joint problem with hybrid beamformers and write the joint antenna selection and hybrid beamformer design problem  as
	\par\noindent\small
	\begin{align}
	& \underset{q_A \in \mathbb{Q}_A}{\maximize} \log_2\bigg|\mathbf{I}_{N_S}\hspace{-3pt} +  \hspace{-3pt}  \frac{\rho}{N_S\sigma_n^2} \big( \mathbf{W}_{BB_{q_A}}^{H}\hspace{-5pt} \mathbf{W}_{RF_{q_A}}^{H}\hspace{-5pt} \mathbf{W}_{RF_{q_A}}\hspace{-5pt} \mathbf{W}_{BB_{q_A}}\hspace{-2pt} \big)^{-1}\hspace{-2pt} \mathbf{W}_{BB_{q_A}}^{H}\nonumber \\
	&\times 			 \mathbf{W}_{RF_{q_A}}^{H} 
	\mathbf{H}_{q_A} \mathbf{F}_{RF_{q_A}}\mathbf{F}_{BB_{q_A}}
	\mathbf{F}_{BB_{q_A}}^{H} \mathbf{F}_{RF_{q_A}}^{H}
	\mathbf{H}_{q_A}^H 
	\mathbf{W}_{RF_{q_A}}\mathbf{W}_{BB_{q_A}} \bigg|  \nonumber \\
	&\subjectto \;\;\; \mathbf{F}_{RF_{q_A}} \in \mathcal{F}_{RF},
	\;\;
	||\mathbf{F}_{RF_{q_A}}\mathbf{F}_{BB_{q_A}}||_\mathcal{F}^2 = N_S, \nonumber\\
	& \;\;\;\;\;\;\;\;\;\;\;\;\;\;\;\;\;\;\;\;	\mathbf{W}_{RF_{q_A}} \in \mathcal{W}_{RF},
	\label{def:RAntennaSelectionHB}
	\end{align}\normalsize
	which requires to solve hybrid beamformer design problem for each $q_A \in \mathbb{Q}_A$. Let us, for now, assume that the hybrid beamformers are estimated as described in the next subsection. Then, the antenna subarray that provides the maximum spectral efficiency is obtained and this subarray is represented by the subarray index $\bar{q}_A$. When the problem in (\ref{def:RAntennaSelectionHB}) is solved for different channel matrices, some of the $\bar{q}_A$ values turn out to be the same for different channel matrices which are similar to each other and the same antenna subarray provides the maximum spectral efficiency for these channel matrices \cite{deepLearningAntennaSelectionElbir}. As a result, the number of subarrays providing the maximum spectral efficiency, say $\bar{Q}_A$, is much less than the number of all subarray configurations, i.e., $\bar{Q}_A \ll Q_A$. In \cite{deepLearningAntennaSelectionElbir}, a similar observation is made for cognitive radar scenario. Therefore, we define another subset of subarray configurations as $\mathbb{A} = \{\mathbb{A}_1,\dots, \mathbb{A}_{\bar{Q}_A}  \},$ which is composed of the subarrays providing maximum spectral efficiency for different channel matrices. In other words, $\mathbb{A}_{\bar{q}_A}$ includes the antenna positions of the subarray obtained by solving (\ref{def:RAntennaSelection}), hence we have $\mathbb{A}\subset \mathbb{S}$.

	For very large number of antennas, say, $N_R > 64$, the computation of all possible antenna subsets is computationally prohibitive and requires very large amount of memory. To tackle this problem, we use close-to-optimum branch-and-bound (BAB) techniques \cite{mimoAntennaSelectionBAB}. To further reduce the complexity, we also partition $\mathbb{S}$ into $B$ non-overlapping blocks as $\mathbb{S}^{(b)} = \mathbb{S}_{N_A(b -1) + 1},\dots, \mathbb{S}_{N_Ab}$ where $N_A$ is the block size and $b = 1,\dots, B$. Then the antenna selection problem is solved for $N_A$ nodes at a times hence less memory is used. In particular, the following strategy is used:
	\begin{enumerate}
		\item For $b = 1$, construct $\mathbb{S}^{(b)}$ and solve (\ref{def:RAntennaSelectionHB}) as $\bar{q}_A^{(b)} = \argmax_{q_A \in \mathbb{S}^{(b)}} R(q_A)$.
		\item For $b > 1$, clear $\mathbb{S}^{(b-1)}$ from the memory and construct $\mathbb{S}^{(b)}$, then  solve (\ref{def:RAntennaSelectionHB}) as $\bar{q}_A^{(b)} = \argmax_{q_A \in \mathbb{S}^{(b)}} R(q_A)$ and obtain the new best subarray index as
		\begin{align}
		\bar{q}_A^{(b)} := \left\{
		\begin{array}{ll}
		\bar{q}_A^{(b)},& \mathrm{if }\; R (\bar{q}_A^{(b-1)}) 
		< R (\bar{q}_A^{(b)})  \\ \bar{q}_A^{(b-1)}, & \mathrm{otherwise}
		\end{array} 
		\right. .
		\end{align}
	\end{enumerate}  
	
	\subsection{Hybrid Beamformer Design}
	\label{sec:RFChainSelection}
	In (\ref{def:RAntennaSelectionHB}), antenna selection is performed by estimating the beamformer weights for each $q_A$. Let $\mathbf{H}_{q_A}\in \mathbb{C}^{N_{RS}\times N_T}$ be the selected channel matrix, then the hybrid design problem can be written as
	\small
	\begin{align}
	&\hspace{-5pt}\underset{\mathbf{F}_{RF} ,\mathbf{F}_{BB},\mathbf{W}_{RF},\mathbf{W}_{BB}}{\maximize}\hspace{-3pt}\log_2\bigg|\mathbf{I}_{N_S} +    \frac{\rho}{N_S\sigma_n^2} \big(\mathbf{W}_{BB}^H\mathbf{W}_{RF}^H \mathbf{W}_{RF}\mathbf{W}_{BB}\big)^{-1}
	\nonumber \\
	& \;\;\;\;\;\;\;\;\;\;\;\;\;
	\times \mathbf{W}_{BB}^H\mathbf{W}_{RF}^H  \mathbf{H}_{{q}_A} \mathbf{F}_{RF}\mathbf{F}_{BB}  \mathbf{F}_{BB}^H\mathbf{F}_{RF}^H \mathbf{H}_{{q}_A}^H \mathbf{W}_{RF}\mathbf{W}_{BB} \bigg|  \nonumber \\
	&\;\subjectto \mathbf{F}_{RF} \in \mathcal{F}_{RF},
	||\mathbf{F}_{RF}\mathbf{F}_{BB}||_\mathcal{F}^2 = N_S,\mathbf{W}_{RF} \in \mathcal{W}_{RF}.
	\label{def:RFSelection}
	\end{align}\normalsize
	Above problem can be written as two decoupled optimization problems to find precoders and combiners separately \cite{mimoRHeath,hybridBFAltMin}. In this case, the Euclidean distance between the unconstrained beamformers and the hybrid beamformers is minimized. In other words, the hybrid precoder design problem can be written as follows
	\begin{align}
	& \underset{\mathbf{F}_{RF}, \mathbf{F}_{BB}}{\minimize}\;\;\; || \mathbf{F}_{q_A}^{\mathrm{opt}} -  \mathbf{F}_{RF}\mathbf{F}_{BB} ||_\mathcal{F}^2 \nonumber \\
	&\subjectto \;\;\mathbf{F}_{RF} \in \mathcal{F}_{RF}.
	\label{def:RFSelection_F}
	\end{align}
	In a similar way, the combiner design problem can be written as
	\begin{align}
	\label{def:RFSelection_W}
	&
	\underset{ {\mathbf{W}}_{RF}, \mathbf{W}_{BB} }{\minimize}\;\;\; || {\mathbf{W}}_{q_A}^{\mathrm{opt}} - {\mathbf{W}}_{RF}\mathbf{W}_{BB} ||_\mathcal{F}^2 \nonumber \\
	&\subjectto \;
	{\mathbf{W}}_{RF} \in{\mathcal{W}}_{RF},		\nonumber \\
	&\;\;\;\;\;\;\mathbf{W}_{BB} = (\mathbf{W}_{RF}^H \boldsymbol{\Lambda}_{{{q}}_A}  \mathbf{W}_{RF})^{-1}(\mathbf{W}_{RF}^H\boldsymbol{\Lambda}_{{{q}}_A}\mathbf{W}_{{{q}}_A}^{\mathrm{opt}}),
	\end{align}
	where $\boldsymbol{\Lambda}_{{{q}}_A} = \frac{\rho}{N_S}\mathbf{H}_{{{q}}_A}\mathbf{F}_{RF}\mathbf{F}_{BB}\mathbf{F}_{BB}^H\mathbf{F}_{RF}^H\mathbf{H}_{{{q}}_A}^H + \sigma_n^2\mathbf{I}_{N_{RS}},$
	denotes the covariance of the array output in (\ref{def:arrayOutputSelected}) corresponding to the selected subarray index ${{{q}}_A}$.		
	
	Above optimization problems in  (\ref{def:RFSelection_F}) and (\ref{def:RFSelection_W}) can be effectively solved by MATLAB-based {Manopt} algorithm  \cite{manopt} via manifold optimization \cite{hybridBFAltMin}\footnote{The proposed antenna and RF chain selection framework can also be applied to the uplink case where the received signal model is $\bar{\mathbf{y}}^{\mathrm{UL}} 
		= \sqrt{\rho}\mathbf{F}_{BB}^{\mathrm{Full}^H} \mathbf{F}_{RF}^{\mathrm{Full}^H} \mathbf{H}^{\mathrm{UL}}\mathbf{W}_{RF}\mathbf{W}_{BB}\mathbf{s} + \mathbf{F}_{BB}^{\mathrm{Full}^H} \mathbf{F}_{RF}^{\mathrm{Full}^H}\mathbf{n},$
		which is obtained by switching the precoders ${\mathbf{F}}_{RF}$, ${\mathbf{F}}_{BB}$ and combiners  ${\mathbf{W}}_{RF}$, ${\mathbf{W}}_{BB}$ in (\ref{def:receivedSignalFullArray}) and the uplink channel matrix is represented by the $N_T\times N_R$ matrix $\mathbf{H}^{\mathrm{UL}} = \mathbf{H}^T$ \cite{mimoHybridLeus1}.}. 
	Once we obtain $\mathbf{F}_{RF},\mathbf{F}_{BB}$ and $\mathbf{W}_{RF},\mathbf{W}_{BB}$, the labels of the output layer of the network can be formed as 
	\begin{align}
	&\mathbf{z} = [\mathrm{vec}^T\{ \angle \mathbf{F}_{RF} \}, \operatorname{Re}\{\mathrm{vec}^T\{\mathbf{F}_{BB}\}\},\operatorname{Im}\{\mathrm{vec}^T\{\mathbf{F}_{BB}\}\}, \nonumber \\
	& \;  \mathrm{vec}^T\{ \angle \mathbf{W}_{RF}\},\operatorname{Re}\{\mathrm{vec}^T\{\mathbf{W}_{BB}\}\},\operatorname{Im}\{\mathrm{vec}^T\{\mathbf{W}_{BB}\}\}  ]^T,
	\end{align}
	which is a $G \times 1$ real-valued vector and $G = N_TN_T^{RF} + N_{RS}N_R^{RF} + 2N_S(N_T^{RF} + N_R^{RF})$.

	\begin{figure*}[t]
		\centering
		{\includegraphics[width=1.0\textwidth]{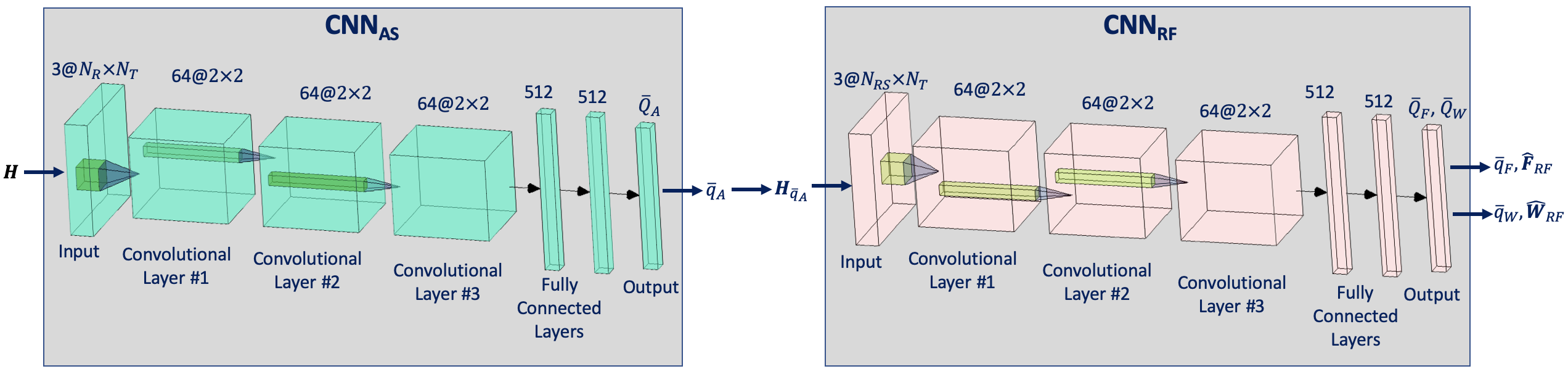}}
		\caption{The proposed CNN architecture for antenna selection and RF beamformer design. }
		\label{figNetwork}
	\end{figure*}

	Even with the above memory-friendly approach, it is still  computationally complex to search all possible antenna subsets in real-time. 
	In order to circumvent this problem, we design a deep learning approach where the network is trained offline with the overhead containing the computation of all possible subarray and RF chain combinations. Then, the trained network can simply be employed as a classification network to select the best antenna subarray and the corresponding RF beamformers for the given channel matrix. We introduce the proposed deep learning technique in the following section.

	\section{Training the DL Network}
	\label{sec:antennaSelectionViaDL}
	The proposed deep network comprises two CNNs (Fig.~\ref{figNetwork}). The first (CNN$_\mathrm{AS}$) accepts the input of channel matrix with the goal to select best antenna subarray $\bar{q}_A$. The second CNN (CNN$_\mathrm{RF}$) takes the input of the subsequent channel matrix with selected rows to choose RF beamformers. For both CNNs, training data are selected from different channel matrix realizations each of which is assigned with the corresponding output classes. 
	
	Let $\mathbf{X}$ be $N_R \times N_T \times 3$ input data of the network with $c=3$ channels. We define the first channel of the input as the absolute value of the imperfect channel matrix $\tilde{\mathbf{H}}$ as $[[\mathbf{X}]_{:,:,1}]_{i,j} = |[\tilde{\mathbf{H}}]_{i,j}|$ where $\tilde{\mathbf{H}} \sim \mathcal{CN}(\mathbf{H}, \boldsymbol{\Gamma})$. $\boldsymbol{\Gamma}\in \mathbb{R}^{N_R \times N_T}$ denotes the variance of AWGN with its $(i,j)$th entry as $\boldsymbol{\Gamma}_{i,j} = \frac{10^{\text{SNR}_\mathrm{TRAIN}/20 }}{[\mathbf{H}]_{i,j}}$ where $\text{SNR}_\mathrm{TRAIN}$ denotes the SNR for the AWGN in the training process. Similarly, the second and the third channels are defined as the real and imaginary parts of $\tilde{\mathbf{H}}$, i.e., $[[\mathbf{X}]_{:,:,2}]_{i,j} = \operatorname{Re}\{[\tilde{\mathbf{H}}]_{i,j}\}$ and $[[\mathbf{X}]_{:,:,3}]_{i,j} = \operatorname{Im}\{[\tilde{\mathbf{H}}]_{i,j}\}$. We generate $NL$ realizations of the channel matrix where $N$ different channel matrices are generated with different user locations and channel gains.  Each channel matrix is generated for $L_\mathrm{c}$ different number of cluster to enrich the training data. In addition, the channel matrix is corrupted by synthetic noise for $L_\mathrm{n}$ realizations  where the element-wise noise is defined by $\text{SNR}_\text{TRAIN}$.
	Hence, the total size of the training input data is $N_R\times N_T\times 3\times NL$ where $L=L_\mathrm{c}L_\mathrm{n}$. For each generated channel matrix, say ${\mathbf{H}}^{(n)}$, the best antenna subarrays with positions $\mathbb{A}_{\bar{q}_A}$ and the best RF beamformers $\mathbf{F}_{RF}, \mathbf{W}_{RF}$ are obtained by solving (\ref{def:RAntennaSelection}), (\ref{def:RFSelection_F}) and (\ref{def:RFSelection_W}) offline. This gives input-output pairs of the training data. The training process of both CNNs is identical except that they have different input dimensions. Algorithm \ref{alg:algorithmTraining} summarizes the steps of training data generation.
	
	\begin{algorithm}[t]
		\footnotesize
		\begin{algorithmic}[1]
			\caption{Training data generation for CNN$_\mathrm{AS}$ and CNN$_\mathrm{RF}$.}
			\Statex {\textbf{Input:} $L_\mathrm{c}$, $L_\mathrm{n}$, $N$, $N_T$, $N_R$, $N_{RS}$, SNR$_{\text{TRAIN}}$}.
			\label{alg:algorithmTraining}
			\Statex {\textbf{Output:} Training data $\mathcal{D}_{\mathrm{AS}}$ and $\mathcal{D}_\mathrm{RF}$.}
			\State Initialize with $t = 1$ for $t = 1,\dots, T = NL$ where $L=L_\mathrm{c}L_\mathrm{n}$.
			\State \textbf{for}  $1 \leq n\leq N$ \textbf{do}
			\State \indent \textbf{for} $1\leq l_\mathrm{c} \leq L_\mathrm{c}$ \textbf{do}
			\State \indent Generate $\mathbf{H}^{(l_\mathrm{c},n)}$ with $N_\mathrm{c} = \mathbb{N}_{\mathrm{cluster}}^{({l_\mathrm{c})}}$.			
			\State  \indent \textbf{for}  $1 \leq l \leq L_\mathrm{n}$ \textbf{do}
			\State  \indent\indent$[\mathbf{H}^{(t)}]_{i,j} \sim \mathcal{CN}([\mathbf{H}^{(l_\mathrm{c},n)}]_{i,j},\sigma_{\text{TRAIN}}^2)$.
			
			\State  \indent\indent\textbf{for}  $1 \leq q_A \leq Q_A$ \textbf{do}
			\State  \indent\indent\indent  $\mathbf{H}_{q_A}^{(t)} = \mathbf{U}_{q_A}^{(t)} \boldsymbol{\Sigma}_{q_A}^{(t)} \mathbf{V}_{q_A}^{(t)^H}$.
			\State  \indent\indent\indent $\mathbf{F}_{{{q}}_A}^{\mathrm{opt}^{(t)}}=\mathbf{V}_{{q}_A}^{(1)^{(t)}}$ .
			\State  \indent\indent\indent $\mathbf{W}_{{{q}}_A}^{\mathrm{opt}^{(t)}} = \bigg( \frac{1}{\rho}\big( \mathbf{F}_{q_A}^{\mathrm{opt}^{(t)^H}}\mathbf{H}_{q_A}^{(t)^H}\mathbf{H}_{q_A}^{(t)}\mathbf{F}_{q_A}^{\mathrm{opt}^{(t)}}  $\par \indent\indent\indent\indent$+ \frac{N_S\sigma_n^2}{\rho}\mathbf{I}_{N_S} \big)^{-1}  \mathbf{F}_{q_A}^{\mathrm{opt}^{(t)^H }}\mathbf{H}_{q_A}^{(t)^H}\bigg)^H$.
			\State  \indent\indent\indent  Use $\mathbf{F}_{{{q}}_A}^{\mathrm{opt}^{(t)}}$, and find ${\mathbf{F}}_{RF_{q_A}}^{(t)}$, ${\mathbf{F}}_{BB_{q_A}}^{(t)}$ by solving (\ref{def:RFSelection_F}).
			\State  \indent\indent\indent Use $\mathbf{W}_{{{q}}_A}^{\mathrm{opt}^{(t)}}$, and find ${\mathbf{W}}_{RF_{q_A}}^{(t)}$, ${\mathbf{W}}_{BB_{q_A}}^{(t)}$ in (\ref{def:RFSelection_W}).
			\State  \indent\indent\indent Compute $R_A^{(t)}(q_A)$ in (\ref{def:R}). 
			\State \indent\indent\textbf{end for}
			\State \indent\indent 
			$  \mathbf{H}_{\bar{q}_A}^{(t)}\leftarrow \bar{q}_A^{(t)} = \arg \max_{q_A} R_A^{(t)}(q_A)$.
			\State  \indent\indent$\mathbf{F}_{RF}^{(t)}\leftarrow {\mathbf{F}}_{RF_{\bar{q}_A}}^{(t)}$, $\mathbf{F}_{BB}^{(t)}\leftarrow {\mathbf{F}}_{BB_{\bar{q}_A}}^{(t)}$.
			\State  \indent\indent$\mathbf{W}_{RF}^{(t)}\leftarrow {\mathbf{W}}_{RF_{\bar{q}_A}}^{(t)}$., $\mathbf{W}_{BB}^{(t)}\leftarrow {\mathbf{W}}_{BB_{\bar{q}_A}}^{(t)}$. 
			\State   \indent\indent$[[\mathbf{X}^{(t)}]_{:,:,1}]_{i,j} = |[\mathbf{H}^{(t)}]_{i,j}|$.
			\State  \indent\indent$[[\mathbf{X}^{(t)}]_{:,:,2}]_{i,j}=\operatorname{Re} \{[\mathbf{H}^{(t)}]_{i,j}\}$ .
			\State  \indent\indent$[[\mathbf{X}^{(t)}]_{:,:,3}]_{i,j} = \operatorname{Im}\{[\mathbf{H}^{(t)}]_{i,j}\}$ $\forall ij$.
			\State \indent \indent $\mathbf{z}^{(t)}  = [\mathrm{vec}^T\{ \angle \mathbf{F}_{RF}^{(t)}  \}, \operatorname{Re}\{\mathrm{vec}^T\{\mathbf{F}_{BB}^{(t)} \}\},\operatorname{Im}\{\mathrm{vec}^T\{\mathbf{F}_{BB}^{(t)} \}\},$ \par \indent \indent$
			\mathrm{vec}^T\{ \angle \mathbf{W}_{RF}^{(t)} \},\operatorname{Re}\{\mathrm{vec}^T\{\mathbf{W}_{BB}^{(t)} \}\},\operatorname{Im}\{\mathrm{vec}^T\{\mathbf{W}_{BB}^{(t)} \}\}  ]^T.$
			\State   \indent\indent Construct the input-output pair $(\mathbf{X}^{(t)}, \bar{q}_A^{(t)})$ for CNN$_\text{AS}$ and \par \indent\indent $(\mathbf{X}_{\bar{q}_A}^{(t)}, \mathbf{z}^{(t)})$ for CNN$_\text{RF}$.
			\State \indent $t \leftarrow t + 1.$
			\State \indent\textbf{end for} 
			\State \indent\textbf{end for} 
			\State \textbf{end for} 
			\State Training data for CNN$_\text{AS}$ and CNN$_\text{RF}$ is obtained from the  collection of the input-output pairs as \par \noindent $\mathcal{D}_{\mathrm{AS}} = ((\mathbf{X}^{(1,1)}, \bar{q}_A^{(1,1)}),\dots,(\mathbf{X}^{(T)}, \bar{q}_A^{(T)})  ) $, \par \noindent  $\mathcal{D}_{\mathrm{RF}} = ((\mathbf{X}_{\bar{q}_A}^{(1,1)}, \mathbf{z}^{(1,1)}),\dots,(\mathbf{X}_{\bar{q}_A}^{(T)}, \mathbf{z}^{(T)})  ).$
		\end{algorithmic}
	\end{algorithm}

	\begin{figure*}[ht]
		\centering
		\subfloat[]{\includegraphics[width=.185\textheight]{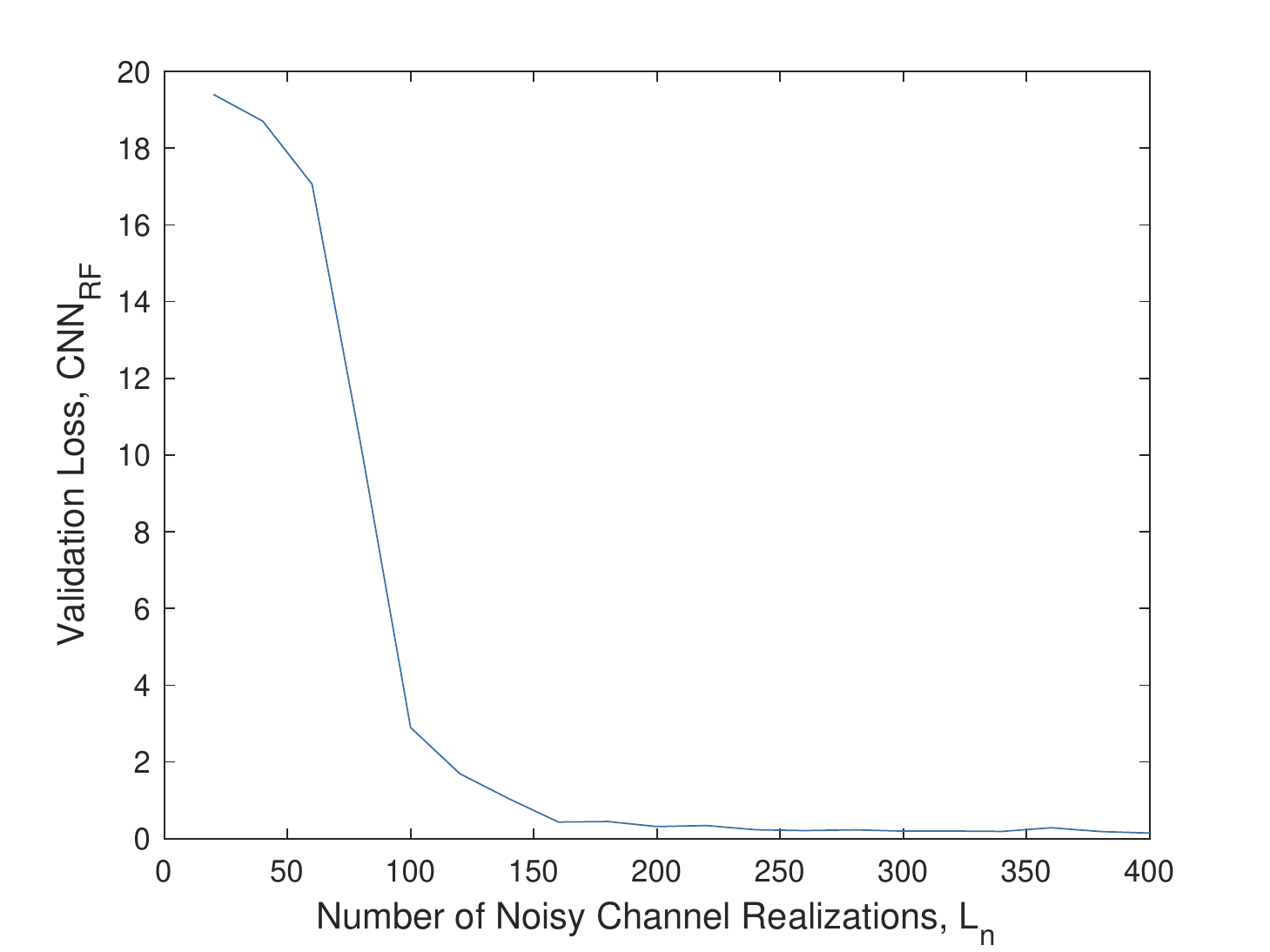}
			\label{figPerformanceOfCNNa} }
		\subfloat[]{\includegraphics[width=.185\textheight]{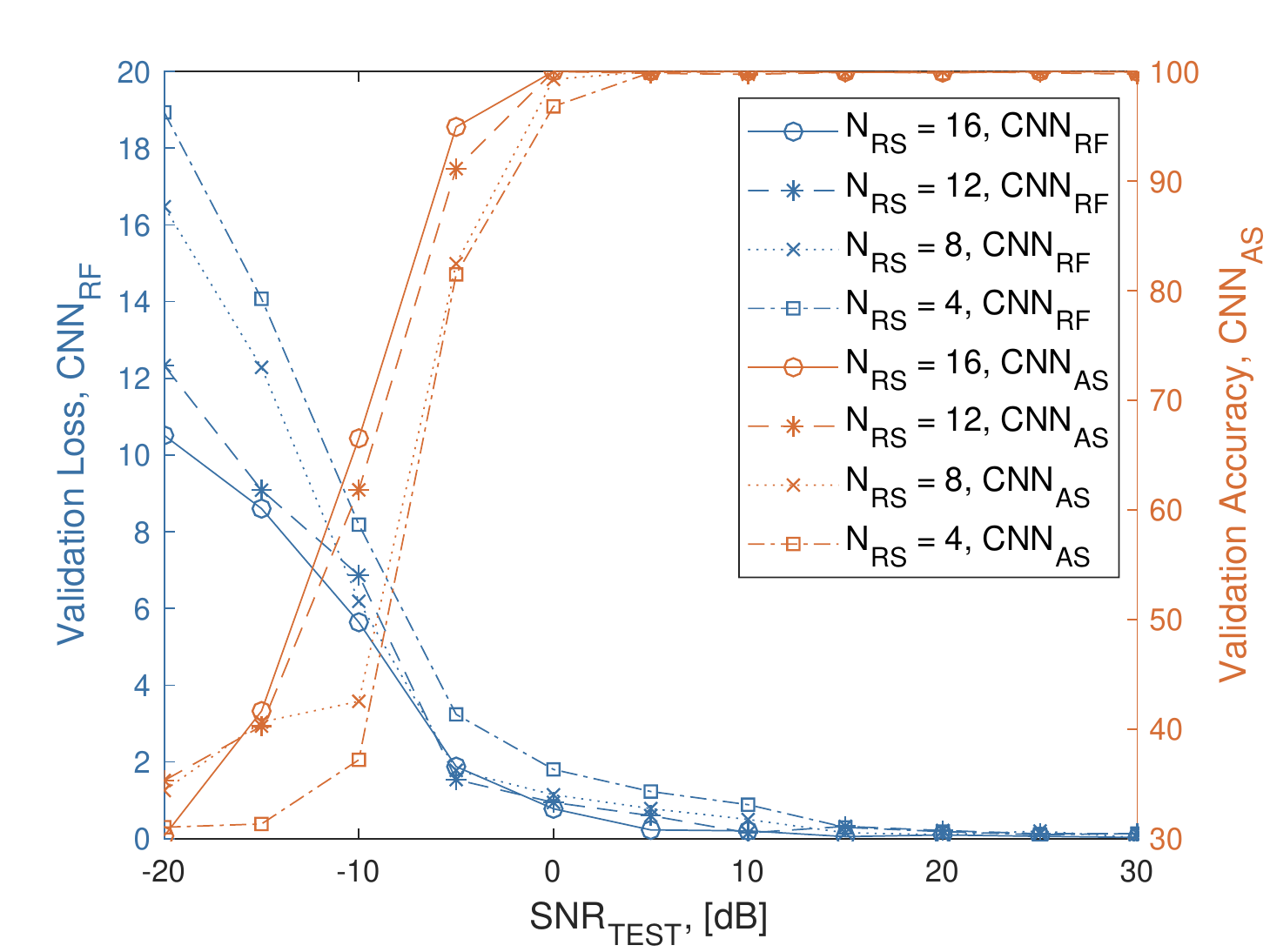}
			\label{figPerformanceOfCNNb} } 
		\subfloat[]{\includegraphics[width=.185\textheight]{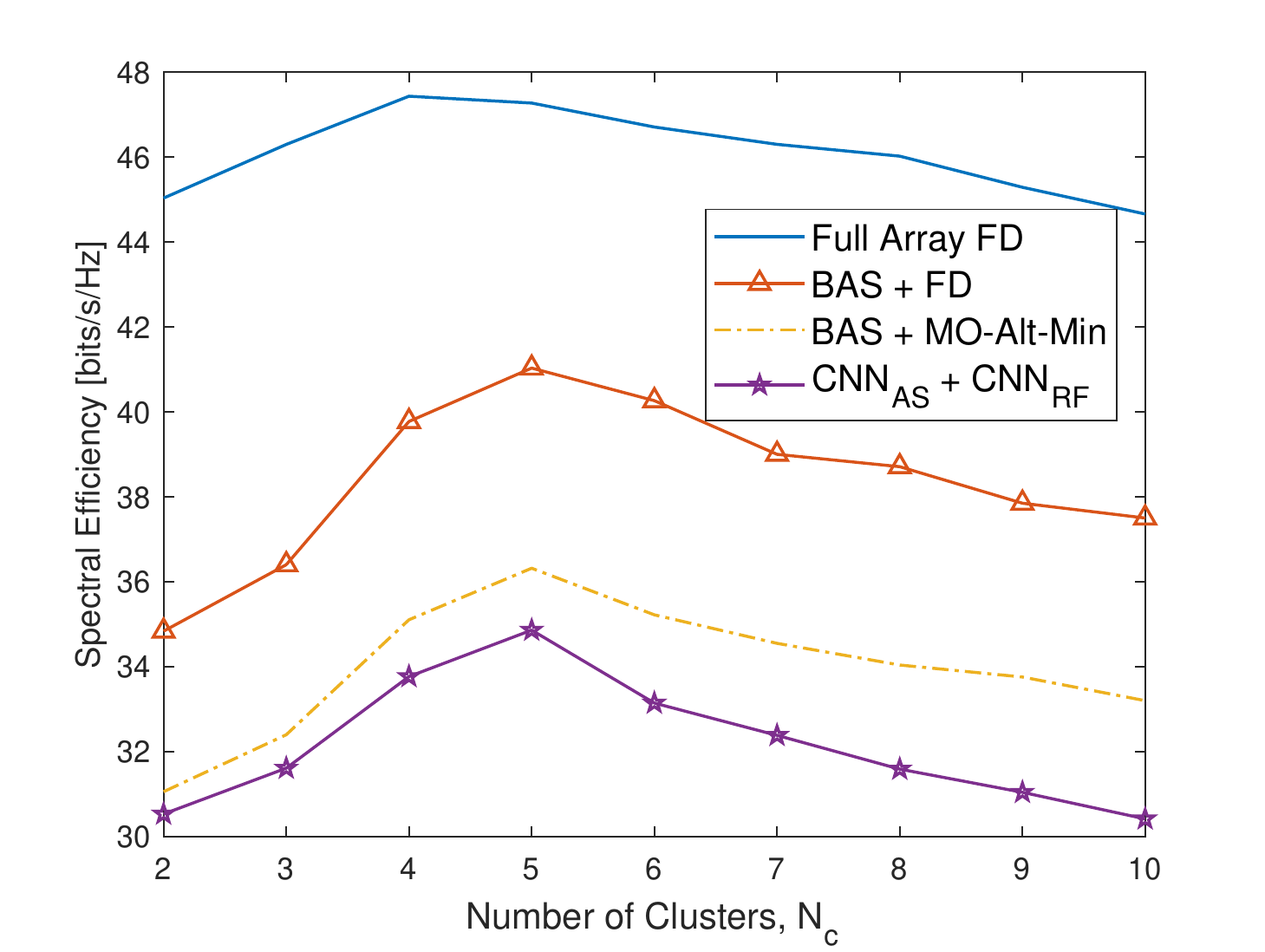}
			\label{figPerformanceOfCNNc} }
		\subfloat[]{\includegraphics[width=.185\textheight]{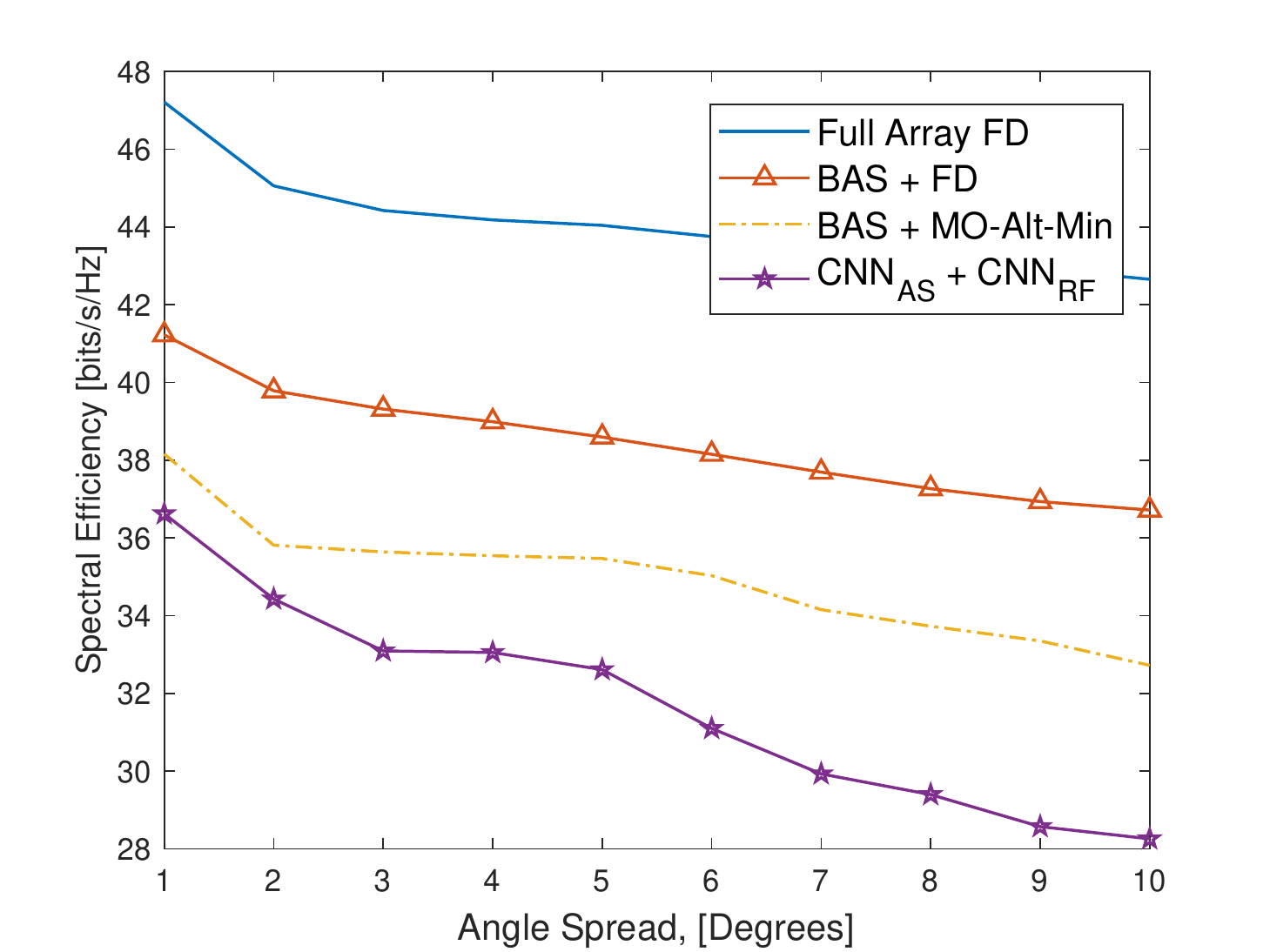}
			\label{figPerformanceOfCNNd} } 
		\caption{Performance of CNN. (a) Validation loss versus number of channel realizations $L_\mathrm{n}$. (b) Validation loss for CNN$_\mathrm{RF}$ and validation accuracy for CNN$_\mathrm{AS}$. (c) Spectral efficiency versus number of clusters in when $\sigma_{\Theta}^2 = 5^\circ$. (d) Spectral efficiency versus angle spread  when $N_\mathrm{c} = 4$. We set $N_T = N_R=64$ and $N_S=4$ for all figures and $N_{RS}=16$ in (a), (c) and (d).}
		\label{figPerformanceOfCNN}
	\end{figure*}
	
	The CNN$_\text{AS}$ accepts the input of size $N_R\times N_T\times 3$ with labels $q_A$ whereas the input of CNN$_\text{RF}$ is of size $N_{RS}\times N_T\times 3$ with label $\mathbf{z}$.  For each CNN, the network is composed of $14$ layers. The first layer is the input layer with appropriate size. The second, fourth and the sixth layers are convolutional layers with $64$ filters of size $2\times 2$. The eight and eleventh layers are fully connected layers with $512$ units. The tenth and thirteenth layers are dropout layers with $50\%$ probability placed after each fully connected layers. There are ReLU (Rectified Linear Unit) layers after each convolutional and fully connected layers where ReLU$(x) = \max(x,0)$. 
	The final layer  of CNN$_\mathrm{AS}$  is the classification layer with size $\bar{Q}_A$ which is the number of subarrays that yield maximum spectral efficiency. In the classification layer, a softmax function is used to obtain the probability distribution of the classes.  The output layer of CNN$_\mathrm{RF}$ is a regression layer of size $G\times 1$. The proposed network is realized in MATLAB on a PC with 768-core GPU. The proposed network architecture is obtained through an optimization analysis to achieve the best performance and less computational cost.

	The use of neural networks in mobile devices is also another issue for practical considerations since a deep neural network is composed of huge number of weights for each layers such as convolutional and fully connected layers \cite{quantizedCNN}. The proposed CNN  framework  should  be  applied  to  the  channel  matrix in  a  mobile  device  to  obtain  hybrid  beamformers  where  the resolution  of  the  network  parameters  is  of  great  importance. In common CNN architectures, convolutional layers have relatively less parameter to be optimized as compared to the fully connected layers which have large number of neurons to be updated in each iteration  \cite{vggRef}. In order to obtain low resolution CNN structure, each of the weights and bias of all  layers are quantized so that the saving in the memory is achieved and the implementation of the CNN is eased.

	To train the proposed CNN structure, $N=100$ different realizations of the channel matrix which is generated for $L=800$  ($L_\mathrm{c} = {4}$ for $N_\mathrm{c} \in \mathbb{N}_\mathrm{cluster} =  \{3,4,5,6\}$ and $L_\mathrm{n} = 200$) noisy realizations with three noise levels, i.e., $\text{SNR}_{\text{TRAIN}}\in \{15,20,25\}$dB.  Hence the total size of the training data is $N_R\times N_T\times3 \times 240000$. In the training process, $70\%$ and $30\%$ of all data generated are selected as the training and validation datasets, respectively. Validation aids in hyperparameter tuning during the training phase to avoid the network simply memorizing the training data rather than learning general features for accurate prediction with new data. The validation data is used to test the performance of the network in the simulations for $J_T=100$ Monte Carlo trials.

	\section{Numerical Experiments}
	\label{sec:Sim}

	We evaluated the performance of the proposed CNN approach via several experiments. In order to prevent the similarity between the test data and the training data we also add synthetic noise to the test data where the SNR in testing is defined similar to $\text{SNR}_\text{TRAIN}$ as $\text{SNR}_\text{TEST} = 20\log_{10}(\frac{|[\mathbf{H}]_{i,j}|^2}{ \boldsymbol{\Gamma}_{i,j} })$. We used the stochastic gradient descent algorithm with momentum \cite{bishop2006pattern} for updating the network parameters with learning rate $0.01$ and mini-batch size of $500$ samples for 50 epochs. As a loss function, we use the negative log-likelihood or cross-entropy loss \cite{deepLearningScience}. We select $N_T^{RF} = N_R^{RF}=4$ for all simulations. For each channel matrix realization, the propagation environment is modeled with $N_\mathrm{c}=4$ clusters and $N_{\mathrm{ray}}=5$ rays for each clusters with the angle spread of  $\sigma_{\Theta}^2=5^{\circ}$ for all transmit and receive azimuth and elevation angles which are uniform randomly selected from the interval $[-60^{\circ},60^{\circ}]$  and $[-20^{\circ},20^{\circ}] $ respectively.
	
	\subsection{Performance of Unquantized CNN}
	\label{sec:SimClassification}
	Figure~\ref{figPerformanceOfCNN} summarizes our assessment of the performance of unquantized CNN. Here, we set $N=100$, $N_T = N_R = 64$ and  $N_S = 4$. Figure~\ref{figPerformanceOfCNNa} shows the validation loss of CNN$_\mathrm{RF}$ against the number of noisy channel realizations $L$. We observe that the loss is satisfactory for $L\geq 150$; in our simulations, we keep $L=200$. Note that this is a common result for different number of channel realizations $N$. In fact, we use $N=100$ in our settings to achieve reasonable network accuracy. To investigate the performance of deep networks against different noise levels in the training data, we demonstrate the validation loss of CNN$_\mathrm{RF}$ and the classification accuracy of CNN$_\mathrm{AS}$ in Fig.~\ref{figPerformanceOfCNNb} for different $N_{RS}$ values. It is clear here that both networks attain satisfactory network accuracy for SNR$_\mathrm{TEST} \geq 0$dB. At low SNR regimes, CNN has poor classification performance due to the deviations between the input and the channel matrices used in the training data. In order to make CNN more robust to noisy inputs, we draw the training data for multiple SNR$_\text{TRAIN}$ levels. Nevertheless, noise in the training data expectedly limits the performance since the network cannot distinguish the input data if it is corrupted too much. This issue is also reported in \cite{deepLearningAntennaSelectionElbir,elbirDL_COMML} for multiple SNR$_\text{TRAIN}$ case.
	
	The channel statistics are important parameters that change in very short time in mm-Wave channels. Hence, in the training stage, we feed the network with channel realizations of different $N_\mathrm{c}$ values. To further investigate the performance with respect to different channel statistics, we compare the algorithms in Fig.~\ref{figPerformanceOfCNNc} and Fig.~\ref{figPerformanceOfCNNd} for different $N_\mathrm{c}$ and $\sigma_{\Theta}^2$, respectively. The proposed CNN framework provides robust performance against different channel statistics. The effectiveness of the proposed techniques can be attributed to training the network with several channel statistics and adding synthetic noise for multiple SNR$_\mathrm{TRAIN}$ values.
	
	As a result, the proposed network provides robust performance against the changes in channel statistics without a need to be re-trained. However, when there is a change in the \textit{full array} system parameters such as $N_T$, $N_R$, $N_{RS}$ and $N_S$, the network does need to be re-trained because these parameters directly dictates the dimensions of the network input and output layers.

	\subsection{Antenna Selection Performance}
	\label{sec:AntennaSelectionPerformance}
\begin{figure}[t]
		\centering
		\subfloat[]{\includegraphics[width=.35\textheight]{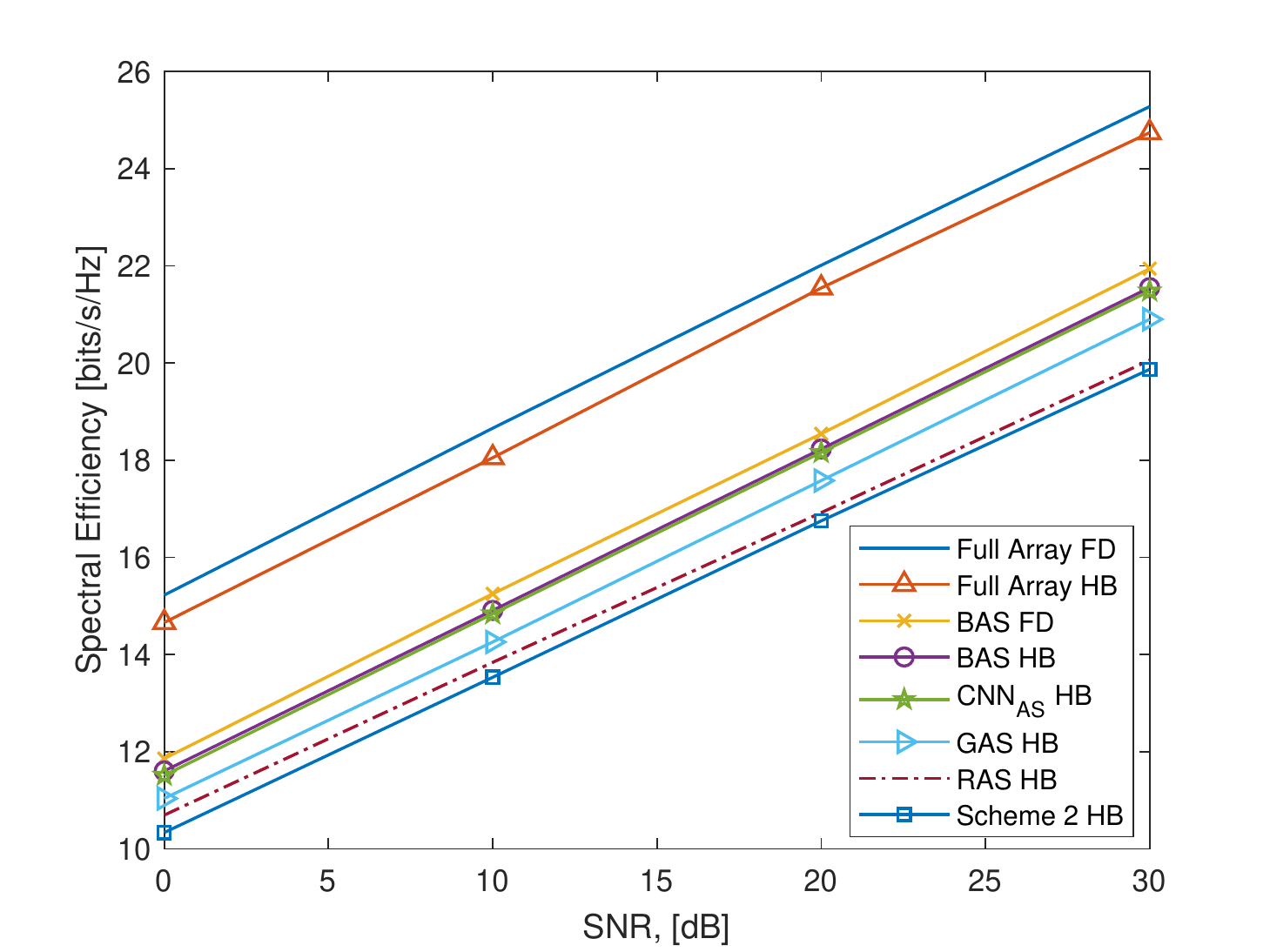} 
			\label{figAntennaSelectionVsSNR}}\\
		\subfloat[]{\includegraphics[width=.35\textheight]{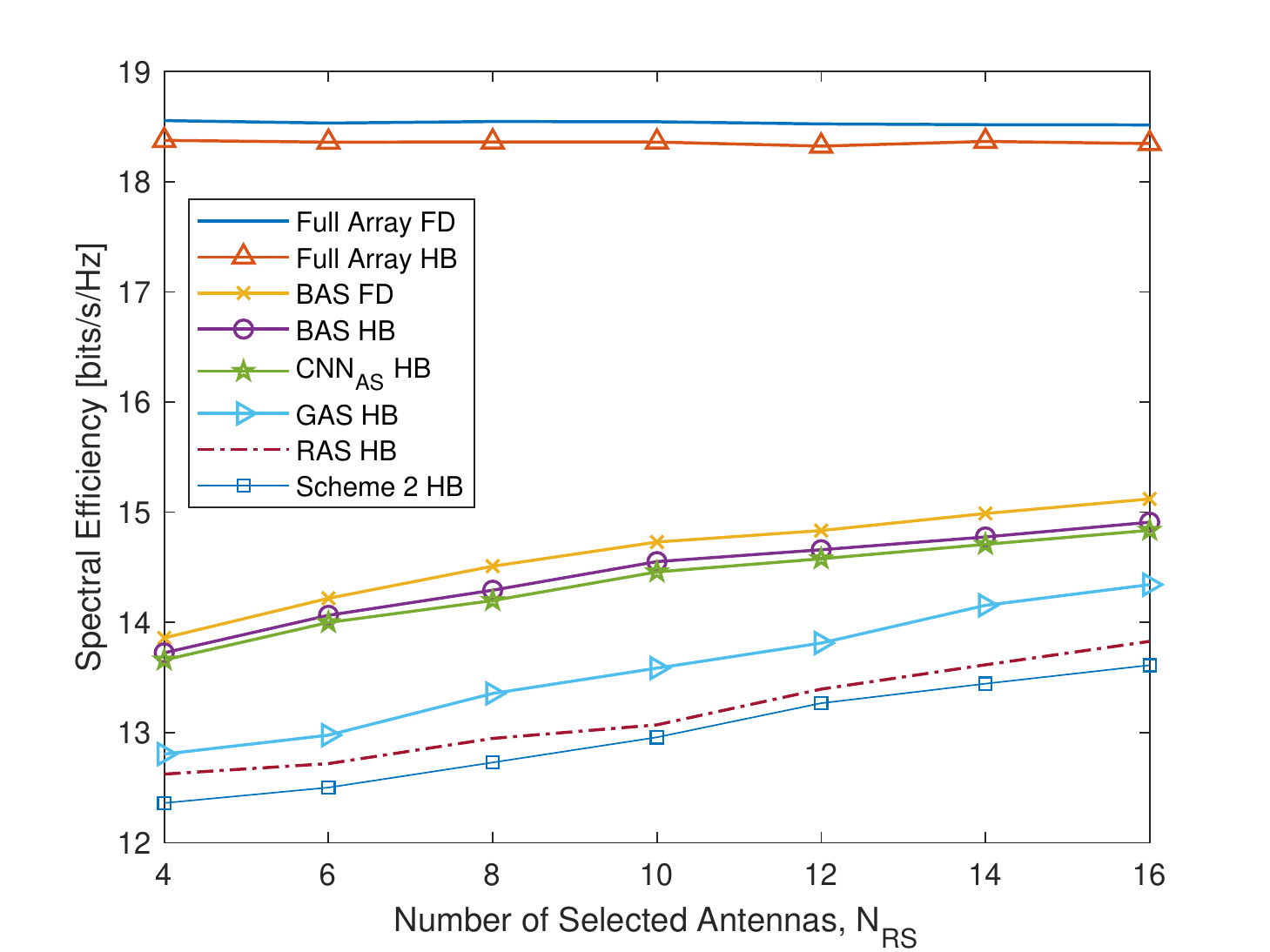}
			\label{figAntennaSelectionVsNumOfRxAntennas} }
		\caption{Antenna selection performance for $N_T = N_R=256$,  $N_{S}=1$: (a) spectral efficiency vs SNR for fixed $N_{RS}=16$ (b) spectral efficiency versus $N_{RS}$.}
		\label{figAntennaSelection}
	\end{figure}

	In this experiment, we present the antenna selection performance for $N_T = N_R = 256$ and $N_{RS} = 16$, and the results are given in Fig.~\ref{figAntennaSelectionVsSNR}. For fair comparison, the hybrid beamforming is performed by  the MO algorithm for all the antenna selection techniques, which are best antenna selection (BAS) with BAB algorithm \cite{mimoAntennaSelectionBAB}, CNN$_\mathrm{AS}$, Greedy-based antenna selection (GAS) \cite{greedyAntennaSelection}, Scheme 1, which is random antenna selection (RAS) and Scheme 2 from Fig.~\ref{figAntennaSelectionSchemes}. Note that BAS, CNN$_\mathrm{AS}$ and GAS aim to optimize the selected antennas and they refer to the selection Scheme 3 demonstrated in Fig.~\ref{figAntennaSelectionSchemes}. The performance of the algorithms is also compared with the full array performance ($N_R = N_{RS} = 256$) which is shown for both fully-digital (FD) and hybrid beamforming (HB). Scheme 1 (RAS) and Scheme 2 have much simpler architectures which do not include optimization and we can see that  Scheme 2 performs the worst since it has the simplest architecture, i.e., selecting the antennas through the absolute values of the entries of the channel matrix. We observe that the performance CNN$_\mathrm{AS}$ is close to BAS, which employs BAB algorithm to yield an optimum solution. Other algorithms/schemes are suboptimal and CNN$_\mathrm{AS}$ outperforms them. 
	The performance of CNN$_\mathrm{AS}$ is superior in comparison with the other algorithms/schemes which provide sub-optimum performance. We also present the performance for different number of selected antennas as shown in Fig.~\ref{figAntennaSelectionVsNumOfRxAntennas} for the same settings and SNR$= 10$ dB. From this figure, we obtain similar observations as in Fig.~\ref{figAntennaSelectionSchemes}, demonstrating the outperformance of CNN$_\mathrm{AS}$.

	\begin{figure}[t]
		\centering
		\subfloat[]{\includegraphics[width=.35\textheight]{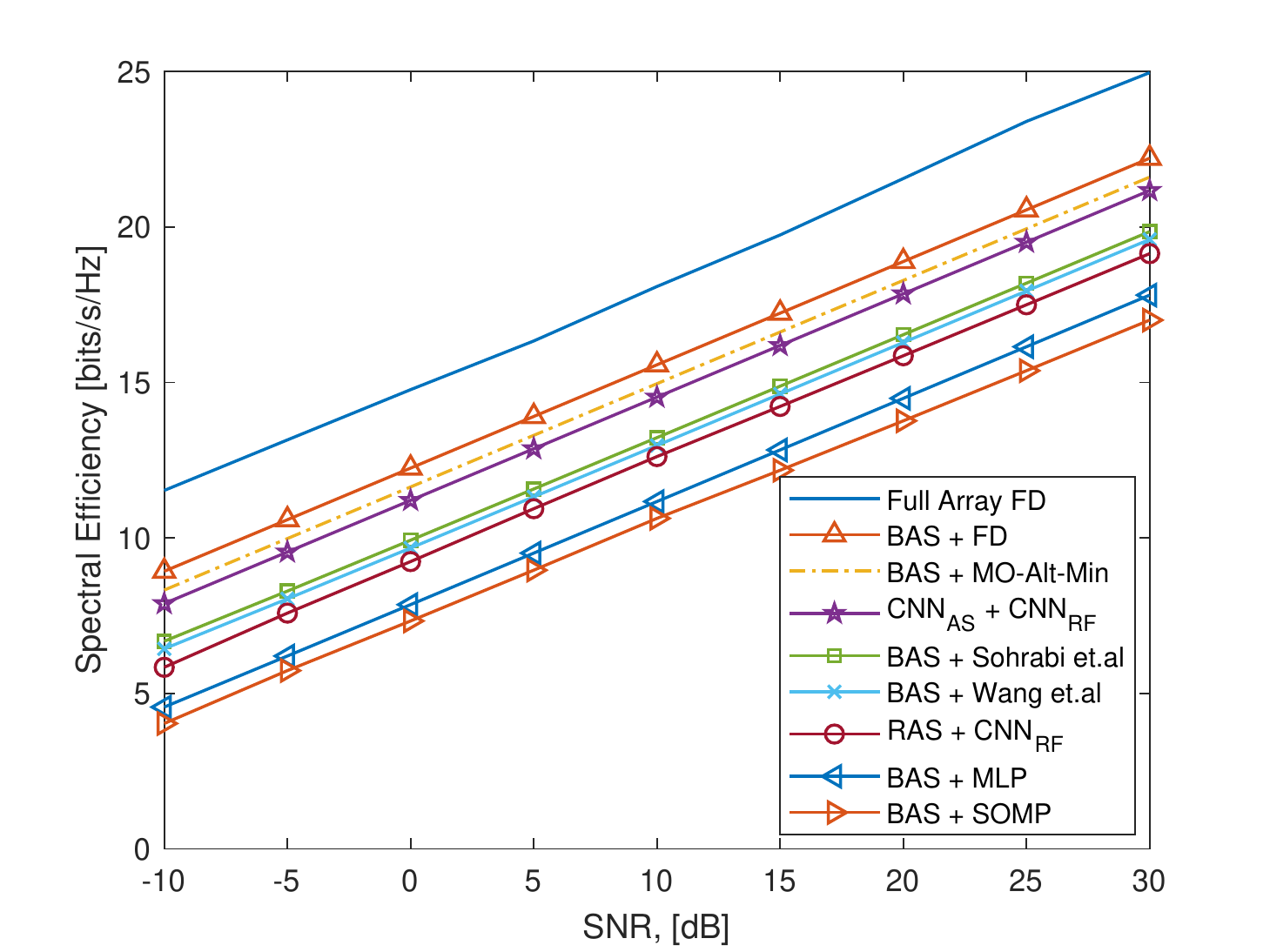}
			\label{figSE_SNRa} } \\
		\subfloat[]{\includegraphics[width=.35\textheight]{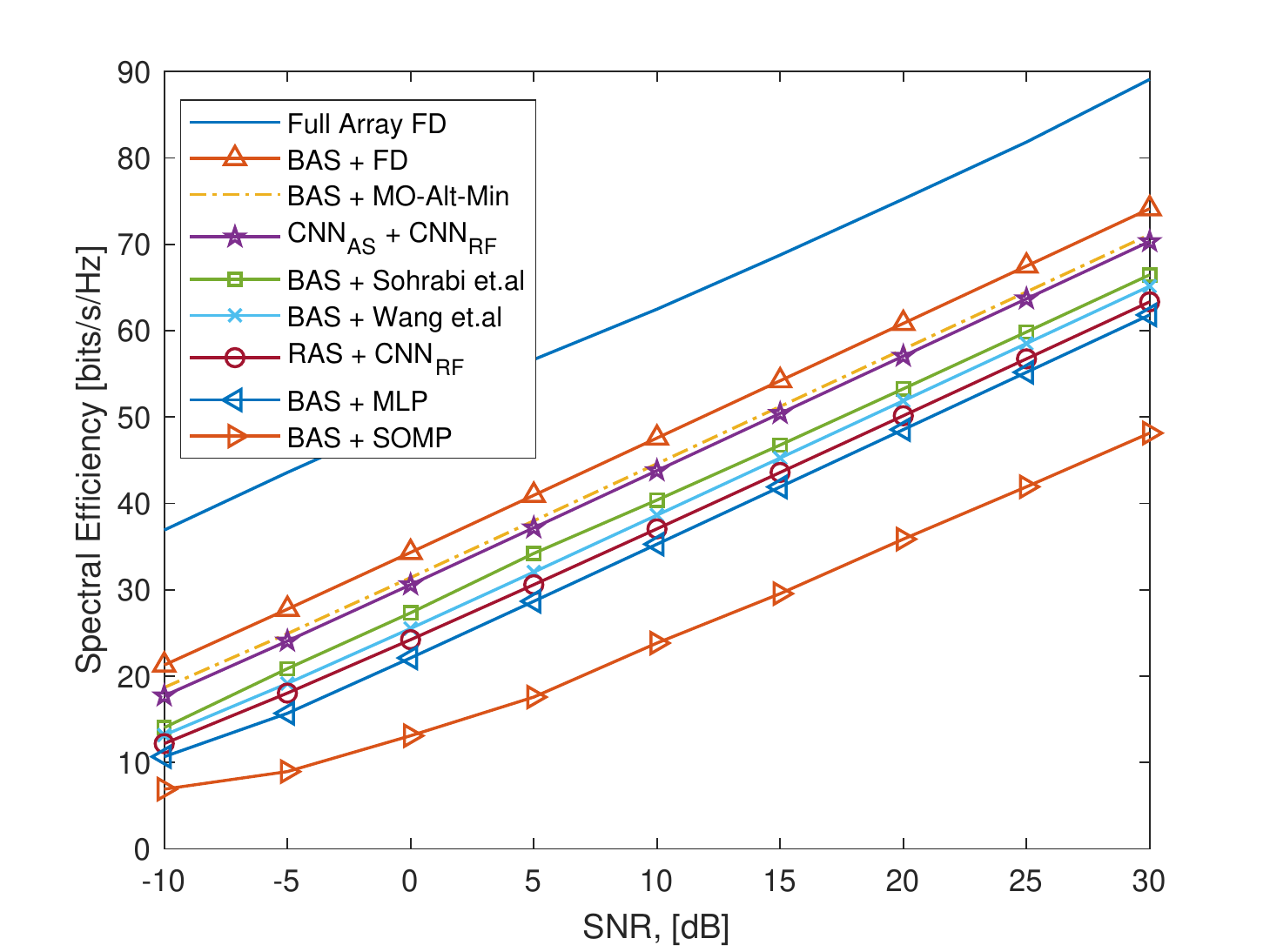} 
			\label{figSE_SNRb}} 
		\caption{Spectral efficiency of various hybrid beamforming algorithms versus SNR for (a) $N_S=1$ and (b) $N_S=4$.}
		\label{figSE_SNR}
	\end{figure}
	
	\begin{figure}[ht]
		\centering
		{\includegraphics[width=.35\textheight]{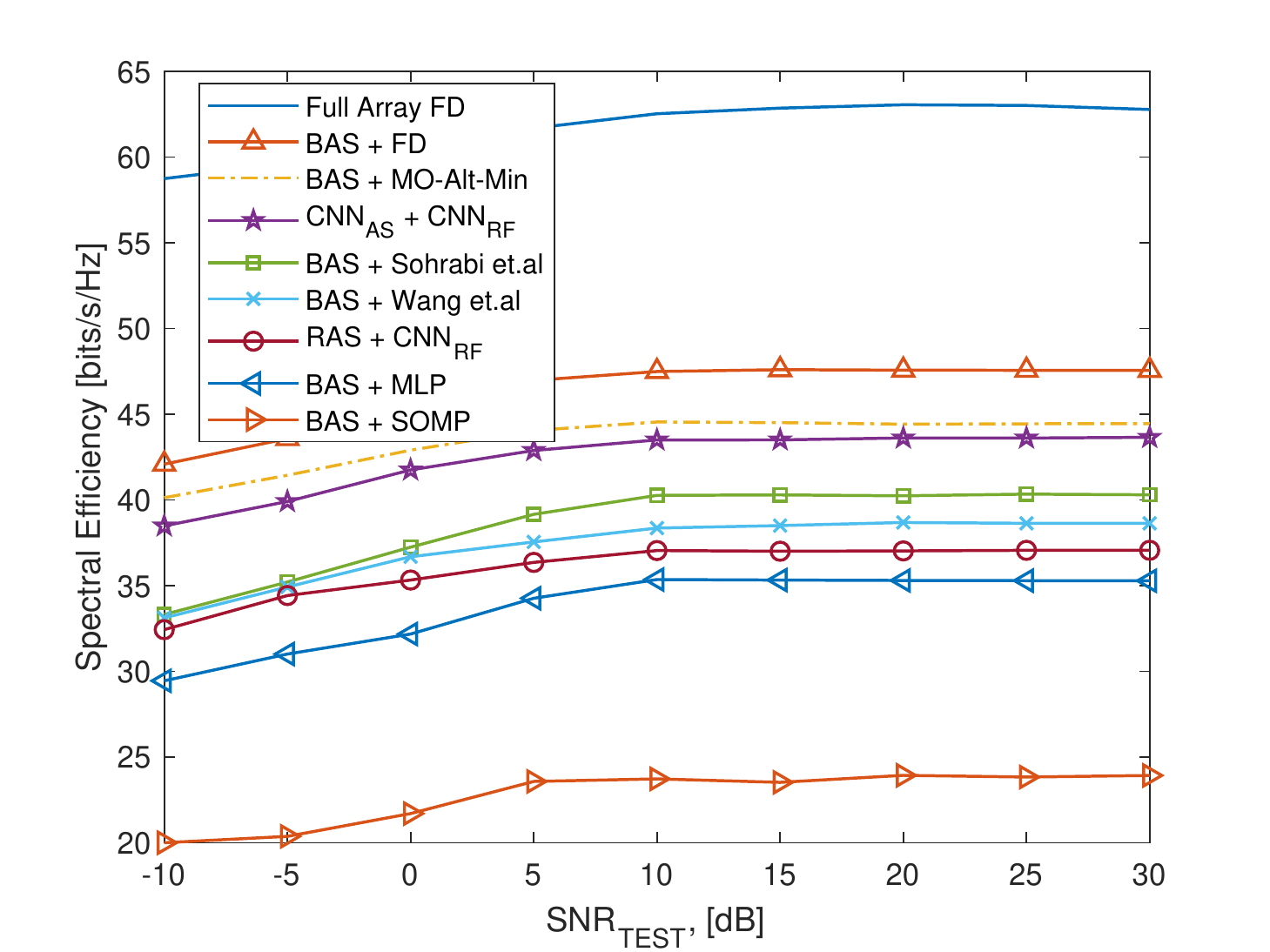} }
		\caption{Performance comparison for corrupted channel data. 
		}
		\label{figCorruptionTest}
	\end{figure}
	
	\subsection{Hybrid Beamforming Performance}
	\label{sec:SpectralEfficiency}
	In this experiment, the spectral efficiency of our CNN-based hybrid beamforming approach  is evaluated by comparison with MO-Alt-Min \cite{hybridBFAltMin}, SOMP \cite{mimoRHeath} as well as the methods in \cite{hybridBFLowRes} (called \textit{Wang~et.al}) and \cite{sohrabiNarrowband} (called \textit{Sohrabi~et.al}).  In addition, we compare our work with the multilayer perceptron (MLP) approach proposed in \cite{mimoDeepPrecoderDesign} where the MLP architecture is fed and trained with the same training data of CNN$_\mathrm{RF}$. The number of antennas are $N_R = N_T = 256$ and $N_{RS}=16$ antennas are selected. For antenna selection, BAS is used for all the hybrid beamforming algorithms except CNN$_\mathrm{RF}$, which is fed by the selected channel matrix obtained  from CNN$_\mathrm{AS}$.  In Fig.~\ref{figSE_SNR}, the spectral efficiency is presented for $N_S=1$ (a) and $N_S=4$ (b) respectively.  We can see that CNN$_{\mathrm{RF}}$ outperforms the other algorithms and it  provides very close performance to MO-Alt-Min which is also used in the labeling process. The performance of CNN$_\mathrm{RF}$ is attributed to the extracting features in the input data and matching the data to labels which are obtained through an exhaustive search algorithm. 
	
	Most of the hybrid beamforming algorithms rely on the perfectness of the channel matrix which is also used in the  design process of beamformers. In order to relax this condition, we use a DL approach and train the deep network with several noisy channel data so that it can obtain robust performance due to the changes in the channel data. In Fig.~\ref{figCorruptionTest}, the robustness analysis is conducted for imperfect channel matrix with respect to  $\text{SNR}_{\mathrm{TEST}}$. The same simulations settings are used with SNR $=10$dB and all algorithms are fed with the imperfect channel matrix. As it is seen, proposed CNN approach performs more robust performance as compared to the other algorithms. The advantage of CNN is that it is trained with several noisy channel data so that it can distinguish the labels for noisy inputs.
	
	We further examine the algorithms on the performance of RF beamformer design. In this respect, we define a cost function to measure how close the algorithms are, to the unconstrained beamformers. We define the error between the unconstrained beamformers  $\textbf{F}^{\mathrm{opt}}$, $\textbf{W}^{\mathrm{opt}}$ and the estimated hybrid beamformers $\hat{\textbf{F}}_{RF}\hat{\textbf{F}}_{BB}$, $\hat{\textbf{W}}_{RF}\hat{\textbf{W}}_{BB}$ as follows
	\begin{align}
	\gamma_F &= ||\textbf{F}^{\mathrm{opt}} - \hat{\textbf{F}}_{RF}\hat{\textbf{F}}_{BB} ||_F/(N_TN_S), \\ \gamma_W &= ||\textbf{W}^{\mathrm{opt}} - \hat{\textbf{W}}_{RF}\hat{\textbf{W}}_{BB} ||_\mathcal{F}/(N_{RS}N_S).
	\end{align}
	Then we present the results in Table~\ref{tableHybridDesginCostNS} for  $N_S=\{1,2,3\}$ and $N_T=N_R=256$, $N_{RS}=16$. It can be seen that CNN-based RF chain selection provides less error for both precoder and combiner design as compared to other algorithm. The performance of the CNN is attributed to aiming the highest spectral efficiency with the selection of the best configuration of beamformers. Note also that the algorithms including our CNN approach achieve less $\gamma_F$ and $\gamma_W$ when $N_S=3$ as compared to $N_S=1$ except SOMP. In contrast,  SOMP performs worse (less $\gamma_F$ and relatively larger $\gamma_W$) when $N_S$ is larger. This explains the performance loss  of SOMP observed in Fig.~\ref{figSE_SNRb} as $N_S$ increases.

	\begin{table}[t]
		\centering
		\begin{threeparttable}[b]
			\caption{Performance Loss For Hybrid beamformer design.}
			\label{tableHybridDesginCostNS}
			\begin{tabular}{ l | l | r | c}
				\hline
				\noalign{\vskip 1pt}
				\multicolumn{2}{c|}{} & $\gamma_F$ & $\gamma_W$  \\[1pt]
				\hline
				\hline
				\noalign{\vskip 1pt}
				& $N_S=1$ & 0.010\cellcolor{black!25}&0.0016 \cellcolor{black!25} \\[1pt]
				\multirow{2}{*}{BAS + MO-Alt-min} & $N_S=2$ &\cellcolor{black!15}0.008 &0.0014\cellcolor{black!15} \\[1pt]
				& $N_S=3$&\cellcolor{black!05}0.005 &\cellcolor{black!5}0.0015 \\[1pt]
				\hline
				\noalign{\vskip 1pt}
				& $N_S=1$ & 0.010\cellcolor{black!25} &0.0060\cellcolor{black!25}  \\[1pt]				\multirow{2}{*}{CNN$_{\mathrm{AS}}$ + CNN$_{\mathrm{RF}}$} 	& $N_S=2$&  0.008\cellcolor{black!15}& 0.0043 \cellcolor{black!15} \\[1pt]
				& $N_S=3$&\cellcolor{black!5} 0.005 & \cellcolor{black!5}0.0022 \\[1pt]
				\hline
				\noalign{\vskip 1pt}
				& $N_S=1$ & 0.032\cellcolor{black!25} & 0.0062\cellcolor{black!25}  \\[1pt]
				\multirow{2}{*}{BAS + \textit{Sohrabi et.al}} & $N_S=2$ &\cellcolor{black!15} 0.019& 0.0034 \cellcolor{black!15}\\[1pt]
				& $N_S=3$ & \cellcolor{black!5}0.008 & \cellcolor{black!5}0.0024 \\[1pt]
				\hline
				\noalign{\vskip 1pt}
				& $N_S=1$ & 0.043\cellcolor{black!25} &\cellcolor{black!25}0.0069 \\[1pt]
				\multirow{2}{*}{BAS + \textit{Wang et.al}} &$N_S=2$& \cellcolor{black!15}0.027& 0.0048 \cellcolor{black!15}\\[1pt]
				& $N_S=3$& \cellcolor{black!5}0.013 &\cellcolor{black!5}0.0028\\[1pt]
				\hline
				\noalign{\vskip 1pt}
				& $N_S=1$ & 0.070\cellcolor{black!25} &\cellcolor{black!5}0.0037  \\[1pt]
				\multirow{2}{*}{BAS + SOMP} & $N_S=2$ & 0.032\cellcolor{black!15} &\cellcolor{black!15}0.0074 \\[1pt]
				& $N_S=3$ &\cellcolor{black!5}0.025 &\cellcolor{black!25} 0.0165  \\[1pt]
				\hline
				\hline
				\noalign{\vskip 1pt}
			\end{tabular}
		\end{threeparttable}
	\end{table}    
	
	\subsection{Binarized and Quantized CNN}
	While considering larger CNN that has more layers and nodes, the associated memory and computational cost could be prohibitive for massive MIMO. This hinders deploying large CNNs in mobile devices, which have limited memory and restricted latency to perform tasks such as online learning and incremental learning. In this context, compressing a deep neural network, that has attracted a lot of attention recently \cite{cheng2017survey}, is highly desirable. In this paper, we adopt a network quantization to compress and thereby accelerate the CNN. This method compresses the original network by reducing the number of bits required to represent each weight of convolutional and fully connected layers. It has been observed that this compression can support both pretrained and trained-from-scratch models, helps in significantly reducing memory usage and speeds up the computations. 
	
	Network quantization in large CNNs can significantly degrade the classification accuracy. We, therefore, investigate the minimum number of bits required to store our proposed twin-CNN network for an acceptable spectral efficiency. Improving the performance of quantized or binarized CNN is an active research area. For example, in the extreme case of binarized or 1-bit CNN, it has been shown that networks trained with back propagation could be resilient to weight distortions introduced by binarization \cite{cheng2017survey}.

\begin{figure}[t]
		\centering
		\subfloat[]{\includegraphics[width=.32\textheight]{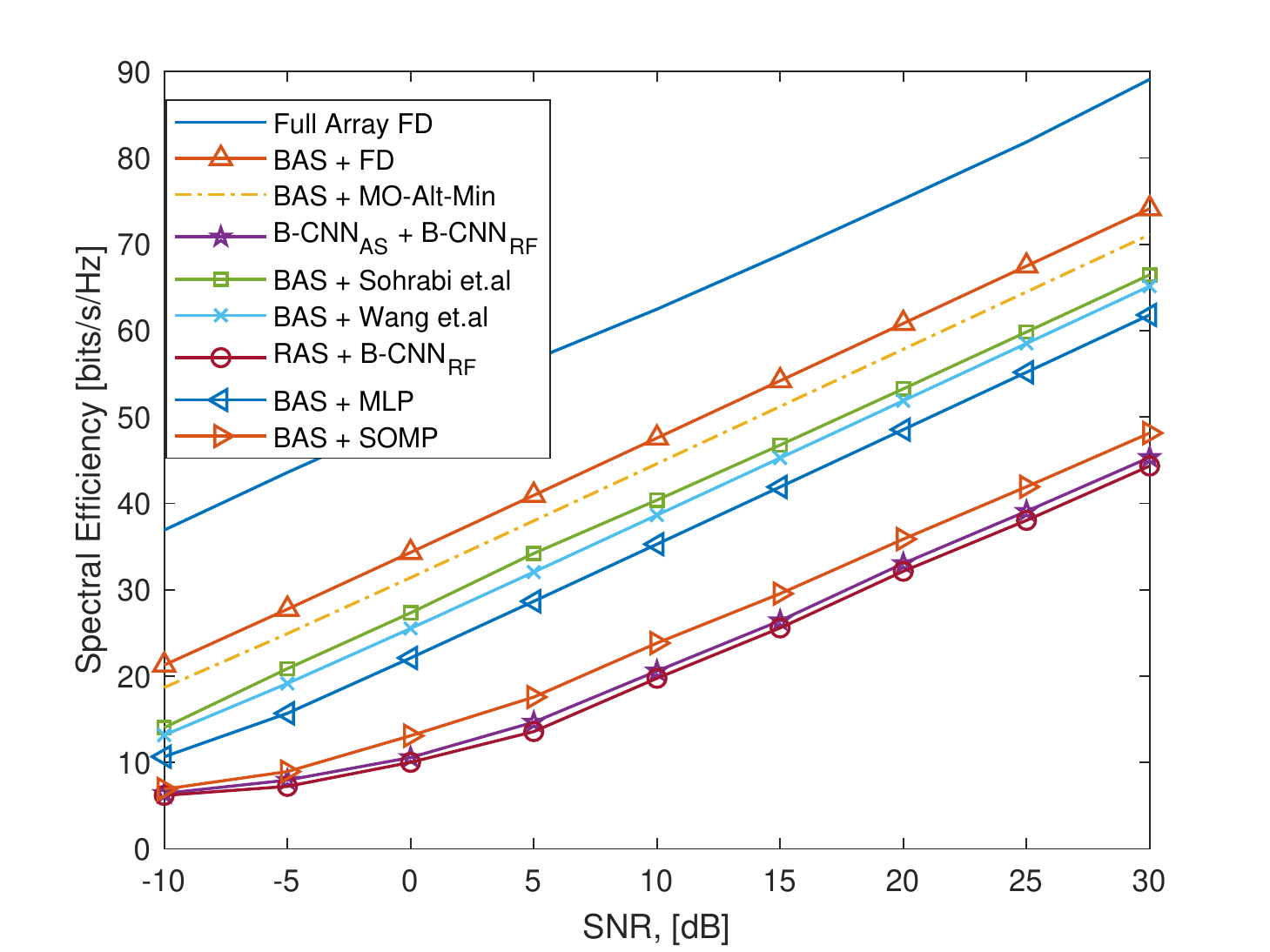}
			\label{figQCNN_SEa} } \\
		\subfloat[]{\includegraphics[width=.32\textheight]{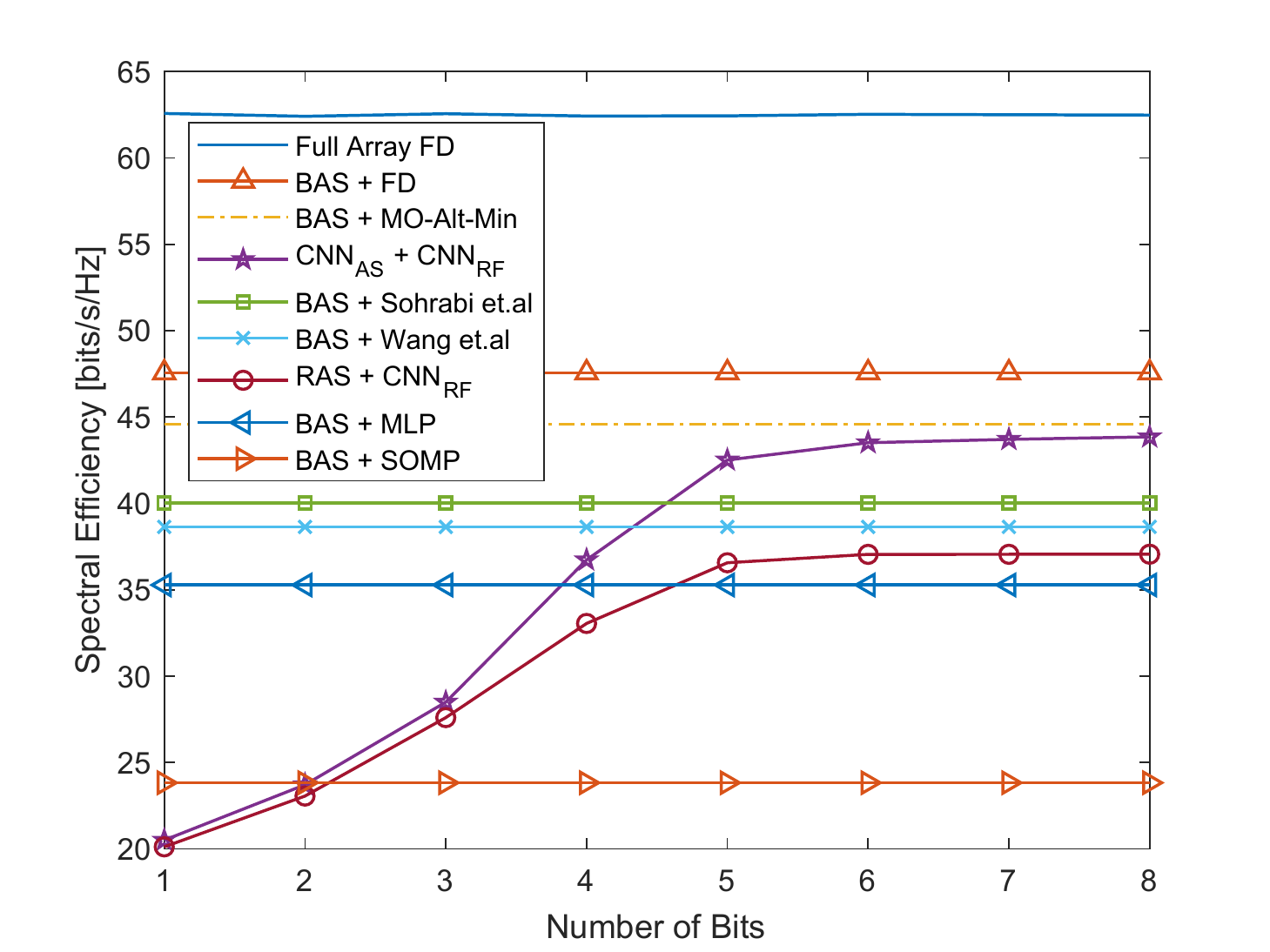}
			\label{figQCNN_SEb} }
		\caption{Performance of  quantized-CNN. Spectral efficiency versus SNR given in (a) and versus number of bits in (b). 
		}
		\label{figQCNN_SE}
	\end{figure}

	We examined the performance of the CNN with quantized weights and biases. In Fig.~\ref{figQCNN_SEa}, we present the performance of binarized-CNN where the parameters of the network are either 0 or 1. CNN performance is poorer than SOMP in case of hybrid beamforming. To further investigate the effect of quantization of network parameters, we demonstrate, in Fig.~\ref{figQCNN_SEb}, the performance of CNN versus number of bits used to quantize the weights and biases of all layers of the CNN. As it is seen, at least 5 bits are required for CNN to attain the best subarray and best RF chain performance.
	

	\subsection{Computational Complexity}
	\label{sec:ComputationalComplexity}
	In this experiment, we measure the computation time of our CNN approach and compare it with other state-of-the-art algorithms. We select  $N_R=64$ and $N_S=4$. The results are given in Table~\ref{tableComputationalComplexity} with respect to the number of BS antennas $N_T$ when $N_{RS}=\{8, 16\}$. As it is seen, our CNN approach enjoys less computation time where the complexity is due to the classification of the input data. The remaining algorithms have relatively high run times and MO-Alt-Min has the highest computation cost however it has better than the remaining algorithms.
	
	\begin{table}[t]
		\caption{Computational Complexity (In Seconds) }
		\label{tableComputationalComplexity}
		\centering
		\begin{tabular}{|c||c|c|c|c|c|}
			\hline
			\multicolumn{6}{c}{ $N_{RS} = 8$} \\
			\hline
			$N_{T}$				 		& CNN					 & MO-AltMin													&\textit{Sohrabi et.al}& \textit{Wang et.al} & SOMP \\
			\hline
			$32$ &    0.003      &      0.241                & 												0.005								& 0.014 &0.028 \\
			\hline
			$128$ &        0.004  &      0.632               &					0.014															 & 0.073 &0.147 \\
			\hline
			$256$ &       0.005   &   1.324                 &						0.041											 & 0.151&0.195 \\
			\hline
		\end{tabular}
		\begin{tabular}{|c||c|c|c|c|c|}
			\hline
			\hline
			\multicolumn{6}{c}{ $N_{RS} = 16$} \\
			\hline
			$N_{T}$				 		& CNN					 & MO-AltMin													&\textit{Sohrabi et.al}& \textit{Wang et.al} & SOMP \\
			\hline
			$32$ &    0.004     &      0.284                 & 			0.008																	& 0.021&0.031 \\
			\hline
			$128$ &        0.010  &      1.171               &					0.027															 & 0.084 &0.112 \\
			\hline
			$256$ &       0.013   &   4.458                  &						0.050													 & 0.172&0.219 \\
			\hline
		\end{tabular}
	\end{table}
	For the sake of completeness, we also calculate the total computation time for the generation of the training data where we select $N_T=N_R=256$, $N_{RS}=16$, $L=N=100$ and use three SNR$_\text{TRAIN}$ levels. In this settings, it takes about two days to generate $256\times 256\times 3\times 30000$ training data. When  $N_T=N_R=25$, $N_{RS}=16$, it takes only 40 minutes for $25\times 25\times 3\times 30000$. In training, the main challenge is not the time but the memory considerations where large antenna arrays yield higher variables of size $Q_A$ which require large memory allocations to save (even temporarily) the results.

	\section{Summary}
	\label{sec:Conclusions}
	We proposed a twin-CNN deep learning approach for joint antenna selection and hybrid beamformer design in mm-Wave communications. 
	Our CNN framework provides significant improvement in the capacity as compared to the conventional beamformer design techniques. This method does not require the precise knowledge of the channel matrix and has significantly better performance than the conventional techniques used in mm-Wave MIMO systems. Instead of computing analog and baseband beamformers, the proposed approach only requires the estimated channel matrix to feed the network and yields the best antenna subarray and analog and baseband beamformers. Hence, it has very low computational complexity. We also investigated the quantized-CNN model when it needs to be applied in a low-memory, low-overhead platform such as a mobile phone. We show that no more than 5 bits are required to save (or access in a cloud-based environment) the CNN in digital form. 
	
	\bibliographystyle{ieeetr}
	\bibliography{main}
	
\end{document}